\documentclass[preprint,5p,times,twocolumn]{elsarticle}

\usepackage{amssymb}
\usepackage{amsmath}

\usepackage{balance}
\usepackage{subcaption}
\usepackage{csquotes}
\usepackage{multirow}
\usepackage{array}
\usepackage{tabularray}
\usepackage[hyphens,spaces,obeyspaces]{url}
\usepackage{hyperref}

\usepackage{xcolor}

\newcommand{\new}[1]{#1}
\newcommand{\edited}[1]{#1}

\usepackage{enumitem}

\journal{Computers \& Graphics}

\begin{document}

\begin{frontmatter}



\title{Scalable Class-Centric Visual Interactive Labeling}


\author[tuw]{Matthias Matt \corref{cor1}}
\cortext[cor1]{Corresponding Author}
\ead{matthias.matt@tuwien.ac.at}
\author[uz,dsi]{Jana Sedlakova}
\author[uz,dsi]{Jürgen Bernard}
\author[fhs]{Matthias Zeppelzauer}
\author[tuw]{Manuela Waldner}

\affiliation[tuw]{organization={Institute of Visual Computing \& Human-Centered Technology, TU Wien},
            city={Vienna},
            postcode={1040}, 
            country={Austria}}
\affiliation[fhs]{organization={Institute of Creative Media Technologies, St. Pölten University of Applied Sciences},
            city={St. Pölten},
            postcode={3100}, 
            country={Austria}}
\affiliation[uz]{organization={Department of Informatics, University of Zurich},
            city={Zurich},
            postcode={8006}, 
            country={Switzerland}}
\affiliation[dsi]{organization={Digital Society Initiative, University of Zurich},
            city={Zurich},
            postcode={8006}, 
            country={Switzerland}}
\begin{abstract}
Large unlabeled datasets demand efficient and scalable data labeling solutions, in particular when the number of instances and classes is large. 
This leads to significant visual scalability challenges and imposes a high cognitive load on the users.
Traditional instance-centric labeling methods, where (single) instances are labeled in each iteration struggle to scale effectively in these scenarios.
To address these challenges, we introduce cVIL, a \textit{Class-Centric Visual Interactive Labeling} methodology designed for interactive visual data labeling. 
By shifting the paradigm from \textit{assigning-classes-to-instances} to \textit{assigning-instances-to-classes}, cVIL reduces labeling effort and enhances efficiency for annotators working with large, complex and class-rich datasets. 
We propose a novel visual analytics labeling interface built on top of the conceptual cVIL workflow, enabling improved scalability over traditional visual labeling. 
In a user study, we demonstrate that cVIL can improve labeling efficiency and user satisfaction over instance-centric interfaces.
The effectiveness of cVIL is further demonstrated through a usage scenario, showcasing its potential to alleviate cognitive load and support experts in managing extensive labeling tasks efficiently.
\end{abstract}



\begin{keyword}


Visual Analytics
Visual-Interactive Data Labeling
Class-Centric Labeling
Property Measures
Interactive Machine Learning
\end{keyword}

\end{frontmatter}



\section{Introduction}

\begin{figure*}[ht!]
    \centering
    \includegraphics[width=0.9\linewidth,trim={0 6.5cm 0 0},clip]{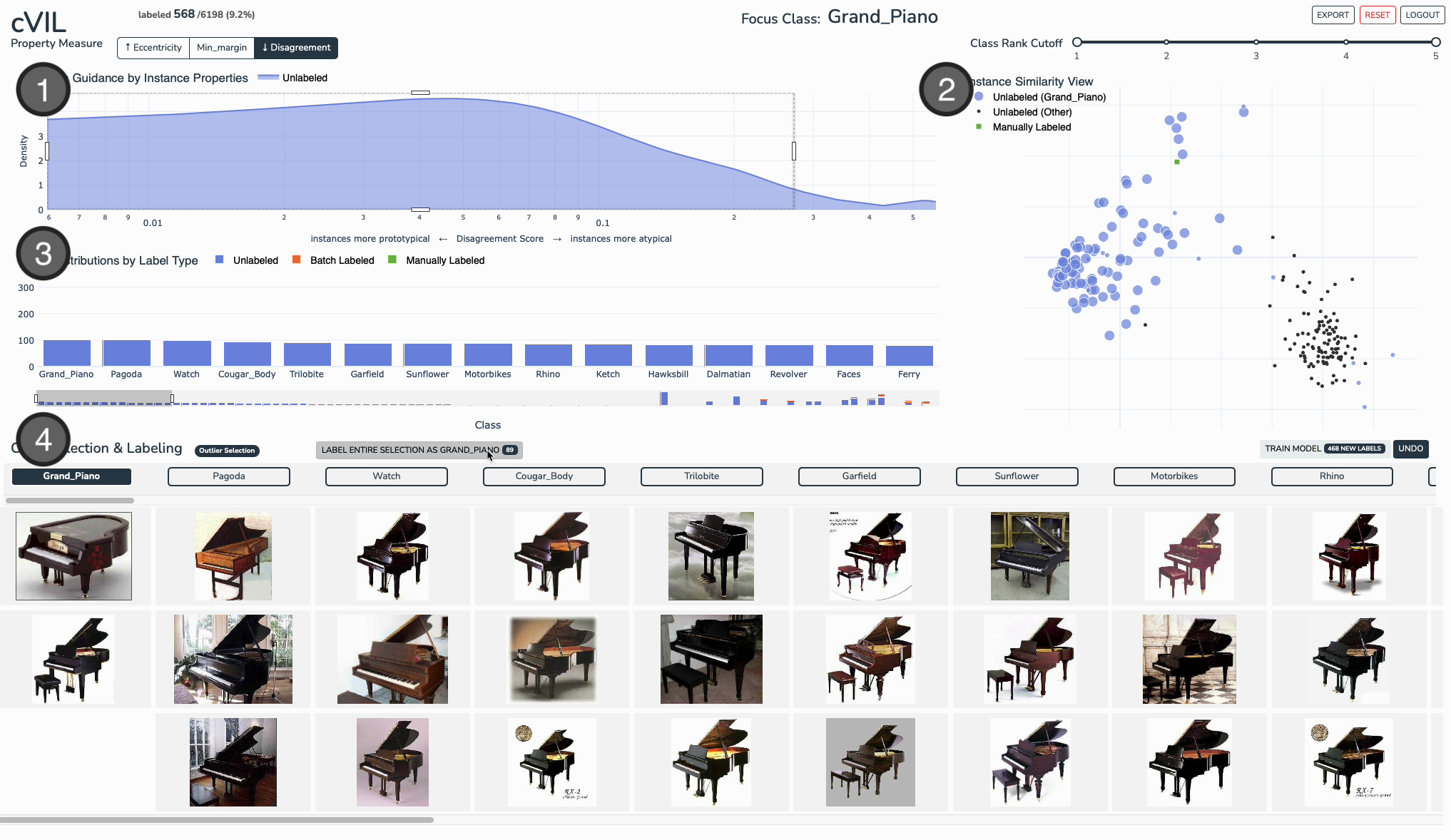}
    \caption{\new{The cVIL prototype contains four main components: The \emph{Instance Property View} (1) displays the distribution of \textit{property measure values}~\cite{Bernard2021ATO} based on all instances predicted for the currently selected focus class. The \emph{Instance Similarity View} (2) shows the instances of the focus class, along with instances that have the focus class predicted as their top-$k$ class (where $k$ is configurable by the user). This plot also visualizes instances that are already labeled within the projection. To get an overview of label distributions, users can utilize the \emph{Class Label View} (3). By clicking on a bar, the focus class is changed to that class. The range selection tool below the main bar chart allows users to increase or decrease the number of classes shown or to move the selection entirely. Once a selection is made in the Instance Property View (1) or Instance Similarity View (2), the selected instances are displayed in the \emph{Instance Labeling View} (4), where users can label either the entire selection or individual instances.}}
    \label{fig:overview_fig}
\end{figure*}

In many scientific and business domains, experts need to analyze and label large amounts of unlabeled data, such as text, images, video sequences, biochemical structures, or measurement data from sensors. 
For many of their downstream tasks, they require \textit{all or at least most instances} of a given dataset to be labeled, i.e., assigned to a known set of classes. 
Thereby, the number of classes might be large, which particularly complicates the labeling task. 
An example is the annotation of topics in social media content by sociologists, where the number of classes (codes, categories, topics) to label can easily reach an order of magnitude of 100 and the number of instances can be huge~\cite{abdelrazek2023topic}. 
Another example is the labeling of bird species in continuous acoustic monitoring recordings by biologists, where the recordings can easily span hundreds of hours containing hundreds of different species~\cite{kahl2023overview}. 
Such large amounts of data, particularly when the number of classes is also large, cannot be inspected and labeled manually in a reasonable time, as the availability of people is typically scarce and expensive.
Computational support is necessary to enable the labeling of such data.

\begin{figure}[t]
     \centering
    \begin{subfigure}[t]{\linewidth}
         \centering
         \includegraphics[width=1.0\linewidth]{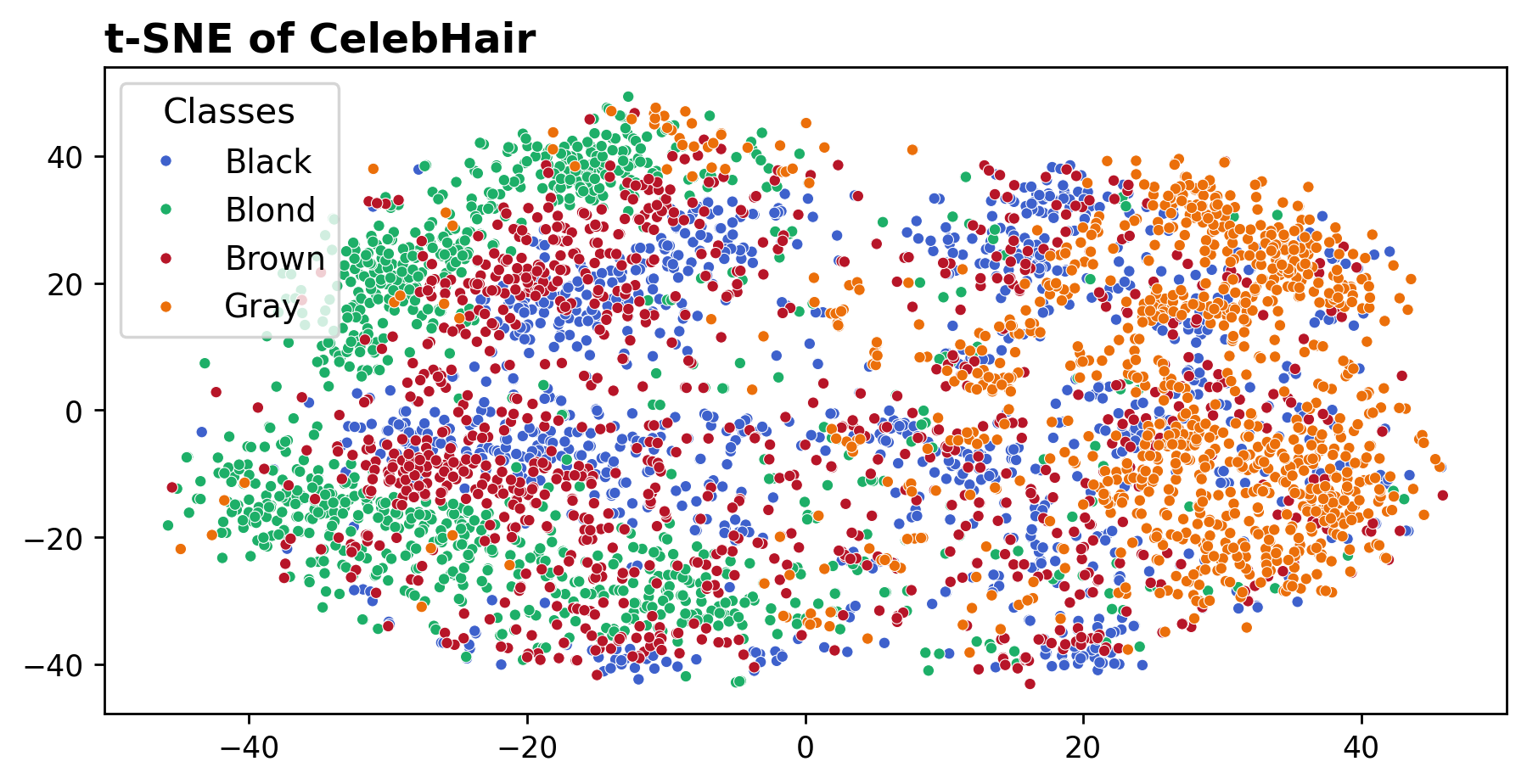}
     \end{subfigure}
    \\
    \begin{subfigure}[t]{\linewidth}
         \centering
         \includegraphics[width=1.0\linewidth]{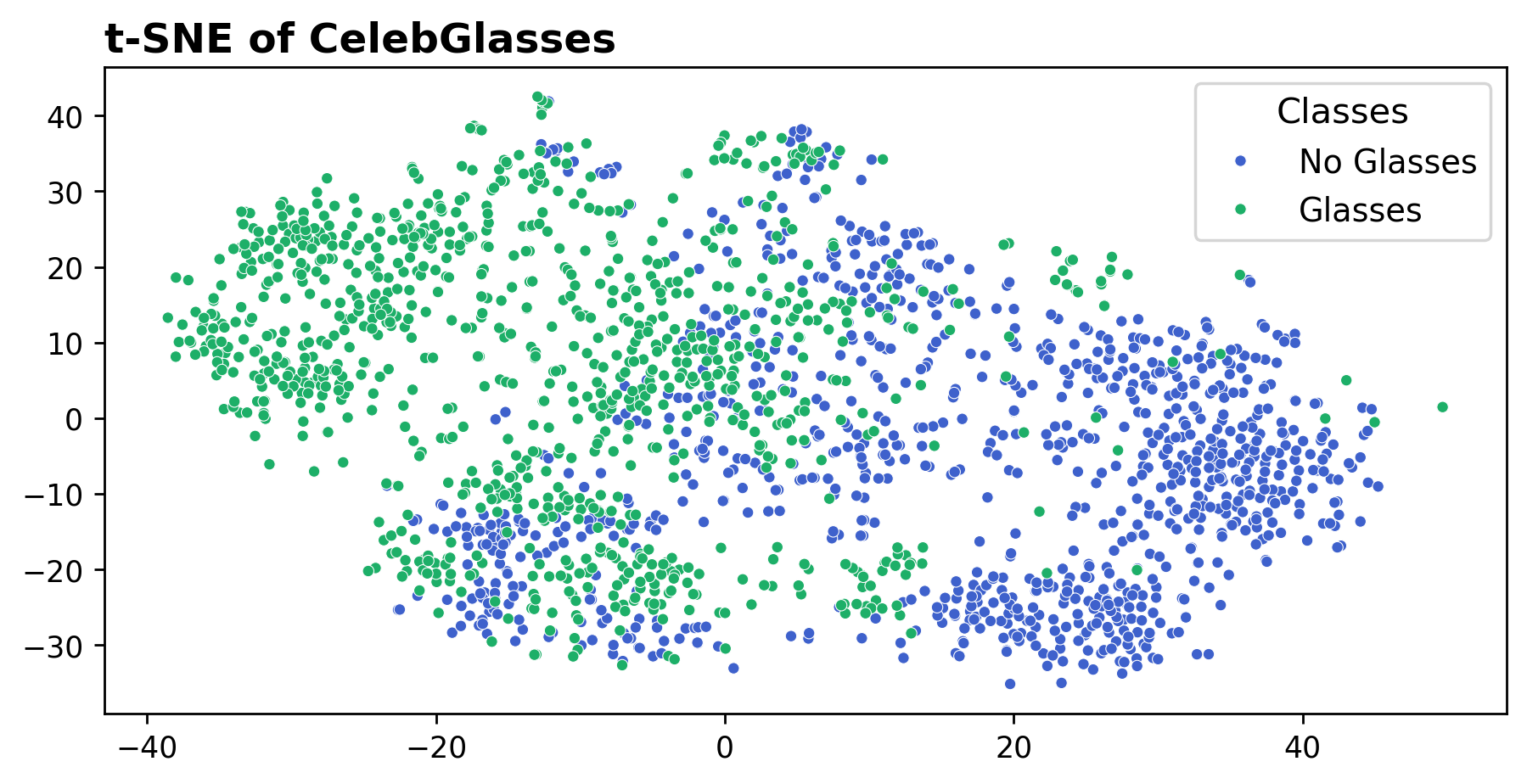}
         \label{fig:tsne_celebg}
     \end{subfigure}
     
    \caption{t-SNE projection of high dimensional embeddings, used for the iVIL approach in our user study. The visualizations are based on embeddings generated by a pre-trained DINO \cite{Caron2021DINO} model, derived from subsets of the CelebA dataset defined by distinct attributes: people's hair color and whether or not they wear glasses. To ensure clarity, we filtered out images classified as having multiple hair colors, only considering those exclusively assignable to one hair color category. The first figure showcases a representative sample of the resulting dataset, featuring 1,000 images for each of the four distinct hair color classes. The second figure shows the projections of the embeddings from people with and without glasses.}
    \label{fig:tsne}
\end{figure}

Reducing the costs for data labeling has been the subject of extensive research.
The overall goal is to obtain labels for all data instances with minimal effort and time investment.
By today, a great majority of data labeling methodologies and approaches are \emph{instance-centric}, i.e., the labeling occurs instance by instance. 
A popular class of instance-centric labeling approaches for interactive labeling is \textit{Active Learning} (AL), which is a model-driven approach. 
In AL, the model selects instances to be labeled by a human. 
The selection of instances is guided by an AL strategy, such as uncertainty sampling \cite{settles2009active, Fu2012ASO} which selects those instances for which the ML model is most unsure about. 
In AL, however, the role of the users is limited to labeling instances autonomously selected by the system, which can become monotonous and frustrating for users \cite{amershi2014power}. 
An alternative instance-centric approach is \textit{Visual Interactive Labeling} (VIL), which is user-driven and enables users to select and label instances through interactive data visualizations. 
In many cases, VIL led to improved results over AL, particularly for less complex annotation tasks \cite{Bernard2018ComparingVL}. 
Typically, these approaches use 2D spatial projections, allowing users to interactively select data instances for labeling. 
To distinguish more easily in the following, we refer to traditional VIL approaches operating on instances as instance-centric VIL, i.e. iVIL. 
A central limitation of iVIL lies in its limited scalability with increasing data complexity, especially with respect to an increasing \emph{number of instances}  and an increasing  \emph{number of classes}: 

    \textbf{Number of instances}: Large datasets challenge the labeling process in different ways. Firstly, labeling at a per-instance granularity quickly becomes impractical in terms of the time required and cognitive load required. Secondly, the scale of the data challenges VIL approaches in that data projection views get increasingly cluttered with growing number of instances, leading to overlaps and occlusions. Figure~\ref{fig:tsne} shows that scatter plots with 1,000 instances per class already lead to considerable clutter. Such crowded visualizations make it inefficient and difficult to select instances for labeling. To tackle the above challenges, a scalable solution is required, which reduces visual complexity and leverages (multi-)instance selection and labeling.
    
    \textbf{Number of classes}: Existing iVIL approaches face significant challenges when the number of classes is high. The predominant method for visual class encoding relies on categorical color schemes, an approach that fails to perform effectively when dealing with more than 12 classes ~\citetext{\citealp[p.~124]{WARE201395}; \citealp{harrower2003colorbrewer}}. Similarly, using a large variety of distinct shapes is difficult for users to differentiate. Additionally, projection-based approaches often fail to achieve adequate visual class separation, as demonstrated in Figure~\ref{fig:tsne}, even with as few as four classes. These limitations can complicate the decision-making process for determining which instances to label and which class to assign. Furthermore, the labeling process with many classes poses a high cognitive burden on the users, requiring users to select from a large set of possible classes for each label-assignment activity. To account for these limitations,  a solution is needed that reduces cognitive load  by simplifying both the visualization and the labeling process.

In this paper, we introduce the \emph{class-centric} VIL workflow (short cVIL), a conceptualization for class-centric data labeling processes. The cVIL workflow covers all phases of the labeling process and incorporates class selection guidance, class-based labeling, but also aspects of \emph{instance-centric} VIL to realize scalability with respect to number of instances and number of classes. cVIL is a novel paradigm for data labeling that shifts the primary focus from instances (as in iVIL) to classes. We design a VA interface based on our previously developed  prototype in~\cite{matt2024cVIL}  that implements the cVIL workflow. While we could show in previous work that our prototype is highly efficient in quickly labeling large amounts of data, its scalability with respect to the number of classes was limited. In this paper, we directly address these challenges. 

\noindent
In summary, our contributions are:
\begin{itemize}[noitemsep,topsep=1pt,parsep=1pt,partopsep=1pt]
    \item the formalization of the class-centric visual interactive labeling (cVIL) workflow to improve scalability, 
    \item the implementation of a novel cVIL labeling interface (Figure~\ref{fig:overview_fig}) with appropriate visualization and interaction design to support the presented workflow, and 
    \item results from an initial user study comparing cVIL with iVIL and a qualitative walk-through demonstrating the potential of cVIL for labeling large data sets with many classes. 
\end{itemize}

\section{Related Work}
\label{sec:relatedWork}

\subsection{Data Labeling Methodologies}
We focus on data labeling methodologies that are instance-centric, i.e., the instance selection and labeling decisions are at the granularity of instances, with a focus on making meaningful choices of next instances, to improve labeling performance and quality.
To the best of our knowledge, no class-centric methodologies have been introduced that allow for human-in-the-loop approaches.

\paragraph{Model-Based Instance Selection}
Active Learning (AL) strategies integrate user knowledge into the learning process, particularly when label information is incomplete, as in semi-supervised learning. 
In AL, a model proactively seeks user feedback (oracle-provided labels) to improve its accuracy \cite{settles2009active}. 
Since user interactions are time-intensive and costly, AL focuses on minimizing queries by targeting information that optimally enhances the model. 
A representative AL workflow is, e.g., depicted in the work of Olsson~\cite{olsson2009}.
To identify the most informative unlabeled instances, various instance selection strategies have been developed and extensively reviewed \cite{settles2009active,olsson2009,tuia2011survey,Wang:2011AL}. 
These strategies are categorized into four main types: (i) uncertainty sampling, (ii) error reduction schemes, (iii) relevance-based selection, and (iv) purely data-centered strategies. 
By employing these methods, AL ensures efficient learning while reducing the dependency on extensive user input.
Reflecting on the state of the art in active learning, Bernard et al.~\cite{Bernard2021ATO} have proposed 15 types of property measures, representing the great majority of AL heuristics in a taxonomic framework.

\paragraph{Human-Based Instance Selection}
VIL-based data labeling methodologies involve users more actively, leveraging visual interfaces for candidate identification. 
Most VIS-based methodologies build upon general process models for visualization \cite{Card1999Foundation,vanWijk2005} and VA \cite{Keim2008VAD,Chen2016}, reflecting abstract data and interaction flows alongside user-driven knowledge generation \cite{SachaSSKEK14}. 
Specific process models for labeling tasks include the work of Höferlin et al. \cite{Hoeferlin2012d}, who introduced an interactive classification method employing Active Learning (AL) strategies, coining ``Interactive Learning'' to emphasize user involvement.
Some proposed processes emphasize user-driven selection of data instances and label assignment based on data-, model-, and user-centered criteria~\cite{Bernard2015AVA}, while other approaches incorporate similarity modeling with user feedback and identify key pitfalls in labeling design~\cite{Bernard2014SimiDef}.
Mamani et al. \cite{Mamani2013} introduced a visualization-assisted method for interacting with data to transform feature spaces.
The VIAL process~\cite{Bernard2018VIALAU} by Bernard et al. builds on these existing works in machine learning and visual-interactive labeling, unifying these approaches into a comprehensive framework that integrates user involvement in labeling tasks. 
By synthesizing prior methods, processes, and strategies, it established the foundation and conceptual groundwork for a series of later implementations, addressing key methodological challenges and enabling innovative applications.

\subsection{Data Labeling Approaches}

\edited{
Following the introduction of VIAL \cite{Bernard2018VIALAU}, several approaches have emerged that implement this process for labeling systems. Earlier works, such as Seifert et al. \cite{Seifert2010UserBasedAL}, visualize the output of a probabilistic classifier for user-based active learning (AL). Grimmeisen et al. \cite{Grimmeisen2022VisGILML} expand on this concept by integrating guidance into the VIAL paradigm, leveraging visual cues in the scatter plot projection for user guidance. Their approach uses the size of glyphs in the scatter plot to indicate information gain and highlight specific instances. Benato et al. \cite{Benato2020SemiAutomaticDA} also utilize a scatter plot to visualize instances, implementing a threshold cutoff where users can divide samples into certain and uncertain instances.
Chegini et ai.~\cite{pacificVAST2019} combine scatter plot visualizations with multivariate data visualizations to support instance-based selection and labeling from different perspectives.
Certain instances are labeled automatically, while uncertain instances require further user labeling, aligning with the batch labeling paradigm in cVIL. Dennig et al. \cite{Dennig2019FDiveLR} introduce FDive, a visual active learning system that allows users to label instances as relevant or irrelevant. This relevance information guides the selection of specific similarity measures, which are used to train a Self-Organizing Map to differentiate between relevant and irrelevant samples.

Another group of approaches focuses on scalability through clustering and automated assistance. Beil and Theisler \cite{Beil2020ClustercleanlabelAI} use clustering to clean data and assign labels to clusters, achieving high scalability in terms of the number of instances. Song \cite{Song2020PersonalizedIC} presents a system for personalized image classification that divides samples into annotation and verification sets using a time-cost optimization approach. Samples in the annotation set are labeled individually by the user, while the verification sets allow for faster labeling, with outliers being labeled manually in subsequent iterations. MorphoCluster \cite{Schrder2020MorphoClusterEA} optimizes label efficiency for large datasets through clustering, grouping similar instances into clusters that are iteratively expanded. This approach uses a ranked list of instances to enable user inspection and selection of similar instances, paralleling the use of property measures in cVIL.
}

\new{
These approaches collectively highlight the importance of visualization, user interaction, and iterative refinement in labeling systems. They demonstrate various methods for enhancing user engagement and decision-making through visualizations and guided interactions. For instance, VIAL and its extensions emphasize the use of visual cues and user strategies to facilitate efficient labeling. Similarly, FDive showcases the effectiveness of relevance-based labeling and the iterative refinement of models. Additionally, clustering methods, such as those employed by Beil and Theisler and MorphoCluster, provide a means to organize and label large datasets efficiently. These approaches collectively underscore the value of visual and interactive methods in improving the efficiency and accuracy of active learning processes.
}

\paragraph{Property Measures}
\edited{
Property measures quantify  specific underlying properties of the model, data, or combination thereof into a single numerical value for each instance. A taxonomy of 15 property measures has been introduced in~\cite{Bernard2021ATO}. In the following, we briefly review algorithms, heuristics, metrics, and measures used in machine learning and visualization research. In cVIL, we leverage the concept of property measures to realize instance selection strategies from model-based AL and user-based VIL approaches.
}

\paragraph{Model-Based Approaches for Property Measures}
Active Learning (AL) strategies often rely on measures to assess data and model properties. 
Common metrics include Manhattan and Euclidean distances for comparing instances or their proximity to class boundaries or cluster centroids~\cite{santini1999,han2011-ch10}. 
While some measures, like cosine similarity, deviate from strict mathematical metrics, others focus on probability distribution comparisons, such as Kullback-Leibler divergence~\cite{kullback1951}, the Kolmogorov-Smirnov test~\cite{kolmogorov1933,smirnov1948} and the Jensen-Shannon divergence~\cite{jensen1906fonctions}. 
Clustering-based AL strategies~\cite{jain2010data} employ Dunn-like index measures~\cite{dunn1974}, Silhouette index~\cite{ROUSSEEUW198753}, Davies-Bouldin measure~\cite{daviesBouldin1979} and Ward's linkage criteria~\cite{ward1963}, see~\cite{Halkidi2002} for an overview. 
Additionally, graph-related metrics~\cite{bonacich1987power}, such as centrality and distances to cluster centroids~\cite{court1960}, are pivotal for assessing the importance of nodes and the centrality of instances. 
These measures are integral to estimating diverse properties, forming a foundational dimension in the design of property measures for AL.

\paragraph{Human-Based Approaches for Property Measures}
In data visualization, property measures are often referred to as visual quality metrics, aiding analysts in detecting patterns like clumpiness or outlierness. 
Wilkinson's Scagnostics measures,~\cite{Wilkinson_scagnostics2005} quantifying visual patterns in scatter plots, are widely recognized, focusing on properties such as outlierness, density, and compactness. 
Extensions of Scagnostics~\cite{dang2014transforming,matute2018skeleton,wang2020improving} have introduced new metrics tailored to specific tasks and visual idioms, such as Magnostics, TimeSeer, and Pixnostics~\cite{behrisch2017magnostics,dang2013timeseer,dasgupta2010pargnostics,lehmann2015visualnostics,schneidewind2006pixnostics,sips2009,Tatu2010a}. 
A growing trend involves modeling human perception to predict how users perceive correlations~\cite{Rensink2010} or cluster patterns in visualizations~\cite{Abbas2019,Aupetit2016,Sedlmair2015}, emphasizing compactness and separation. 
While these measures stem from statistical and perceptual modeling, they align with our focus on human strategies, specifically targeting instance selection strategies to refine property measures in both visualization and machine learning contexts.

Labeling systems provide support across various disciplines, with most focusing on images from different domains. These systems often emphasize either instance-centric or cluster-centric approaches. Pure instance-centric methods, however, can struggle with scalability as the number of instances or classes grows, leading to visual clutter and increased cognitive load, particularly when class distributions overlap. Cluster-based approaches face similar limitations, as numerous class interactions can result in unstable and inaccurate clusterings. User-based approaches generally offer better usability than strictly model-based ones. Property measures enable the integration of model-based ranking with user-based interactions through quantifiable properties, allowing for the inclusion of visual quality metrics and the incorporation of common human labeling strategies into a class-based approach via quantifiable measures. We have identified the need for high scalability in terms of the number of instances and classes as a limitation of previous approaches and aim to improve usability in these scenarios. Our class-based approach introduces new interaction techniques that address common labeling system problems, reducing complexity and enabling new use cases and usage patterns.

\section{Class-Centric Visual Interactive Labeling}
\label{sec:cVIL}

\subsection{cVIL Methodology and Labeling Complexity}

We introduce Class-centric Visual Interactive Labeling (cVIL), a method for (visual) data labeling that shifts the focus from instance-centric labeling to class-centric labeling.
Instead of assigning a label to a focused instance, users find instances belonging to a  class in focus. 
The labeling paradigm changes from ``which class should be assigned to a given instance?'' to ``which instances belong to a specific class?''.
This approach offers several advantages.

\new{
The class-centric focus reduces the cognitive load during labeling because the problem that needs to be solved is just to decide if a certain instance (or several instances) belongs to the focus class. In contrast to instance-based labeling, all other classes can be neglected, making the decision easier. The label assignment problem simplifies to a binary decision: determining whether instances should be assigned to the focus class or not. The cVIL paradigm thereby contrasts with traditional labeling approaches, where users have to choose between multiple labels for a single instance in focus, which becomes increasingly tedious with a growing number of classes. 
}

Methodologically, the focus on  classes motivates multi-instance labeling operations, further increasing labeling efficacy.
Visualization-wise, focusing only on the subset of instances predicted for the focus class, cVIL can reduce clutter in visual labeling interfaces and enables more focused and efficient user interactions, including interactive visual techniques for multi-instance selection and labeling.

\new{The following formalization of the labeling complexity points out the differences between iVIL and cVIL.  
In the case of iVIL, at each step, a user considers instances $i \in S$ from a selection $S \subseteq I$, the whole set of instances $I$, and decides which class label $c \in C$ to assign to it. Here, the user has to first consider $n := |I|$ instances and find a suitable subset $S$ where in the best case this subset contains only instances from a single class which has to be identified from all possible classes $m := |C|$, where $m < n$. To introduce some notation describing the size of the set of instances or classes the user has to work with at a specific stage, we can write this as:
$$ n \rightarrow_{S} m $$
The subscript of the arrow denotes which set of instances the user works with. So the user makes a selection $S$ from the whole set of instances of size $n$ and then has to assign that selection to one of $m$ classes. 


For a single labeling step in cVIL, the user first selects a class $c \in C$ which partitions the instances into a smaller subset $I_c \subseteq I$, which contains $n_c := |I_c|$ instances. With appropriate guidance from the cVIL interface, as we will discuss in the subsequent sections, the user should then be able to select a large subset of instances which belongs to the selected class and label it. In our notation this can be written as:
$$ m \rightarrow_{I_C} n_c \rightarrow_{S} 1 $$


Consider a set of instances with an equal number of instances $n_c$ per class so that $n = \sum_{c \in C} n_c = n_c m$.
In the best case scenario, using iVIL, the interface helps the user to identify $m$ subsets of instances, where each instance in a subset belongs to the same class, and each subset is associated with a different class. In the best case scenario, labeling each of these $m$ subsets yields sufficient model accuracy to conclude the labeling process. In total, the user thereby has to perform $m$ instance subset selections and confirm that the $n_c$ samples within the subset are correct. Deciding between one of the $m$ classes to label the selection can be trivial when an underlying classifier is trained similarly to cVIL, leading to $O(m n_c)$. 

In the best case scenario, using cVIL, the interface supports the user to visit all the $m$ classes in the most effective order so that they can easily identify, for each class, one subset of instances predicted for the selected class and confirm the prediction (i.e., make one binary decision for the single subset). Again, the best case scenario is that one subset selection per class is sufficient for sufficiently training the model. In total, the user has to visit $m$ classes and confirm the correctness of $n_c$ samples, also resulting in $O(m n_c)$.

For the worst case scenario, each instance has to be assigned individually to the correct class. In iVIL, the user has to consider $n$ instances and assign them to $m$ classes individually, requiring $O(nm)$ steps. The same worst case scenario in cVIL leads to the same number of steps as for each of the $m$ classes, each sample within the class has to be assigned individually to the correct class, requiring $O(m^2 n_c)$. Since $n=n_cm$ this is also $O(nm)$.

However, on average, a class will contain clusters with more than one instance per class that can be easily labeled with a single selection. Assume this applies to 10\% of samples ($\frac{n_c}{10}$ samples) within each class.
In cVIL, this remains somewhat efficient, assuming the classifier is accurate enough. The user selects a class, examines the samples with the smallest property measure values or a well-formed cluster within the class and labels these samples. Within each of the $m$ classes, the user only has to consider a subset of the $n_c$ instances. This requires $O(mn_c)$. Since $n=n_cm$ this results in $O(n)$ steps.

In contrast, iVIL requires more effort in this scenario. It is reasonable to assume that the user needs to examine a small subset of the data (consider 10\% of the full dataset or $\frac{n}{10}$ instances) to either identify an appropriate class to label or find an appropriate cluster for the pre-determined class. This process must be repeated for each of the $m$ classes, resulting in a total effort of $O(m \frac{n}{10})$ or $O(n * m)$ to label the same 10\% of the total number of samples.
By reducing the scope of the search from the whole dataset in the case of iVIL to within predicted classes, cVIL can increase labeling efficiency.


}


\subsection{The cVIL Labeling Workflow and Tasks} \label{sec:workfow}

 \begin{figure}
  \centering
  \includegraphics[width=1 \linewidth]{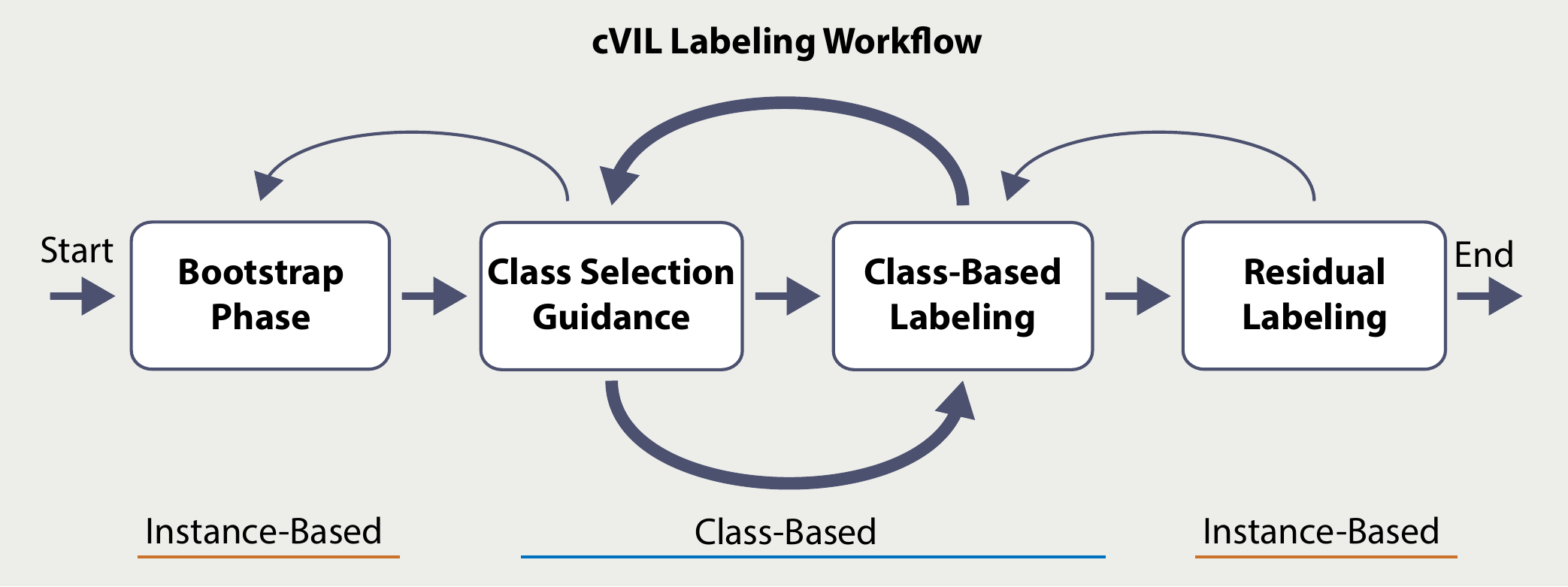}
  \caption{The conceptual cVIL workflow consists of four phases. The process is bounded by a Bootstrap phase and a Residual Labeling phase, both with a traditional instance-centricc focus. In contrast, the core part of the workflow introduces a class-centric focus. In two highly iterative phases, users find means to identify most promising classes to be labeled next, possibly with computational support (Class Selection Guidance), and conduct the Class-Based Labeling.}
  \label{fig:workflow}
\end{figure}  

The cVIL workflow is illustrated in  Figure~\ref{fig:workflow}. It divides the labeling process into four subsequent phases.
In the initial \textbf{Bootstrap} phase, users address the cold start problem by collecting initial sample instances for each class.
With at least one instance labeled per class, supervised machine learning support can be added to the interactive, iterative, and incremental process.
The core cVIL workflow includes two cyclically connected  phases: the \textbf{Class Selection Guidance} phase, where the system identifies the next class to focus on by human-machine collaboration and the \textbf{Class-Based Labeling} phase, where users label multiple recommended instances per class.
The process concludes with the \textbf{Residual Labeling} phase, where users shift back to an instance-based focus to label remaining (unsolved or out-of-distribution) instances. 
In the following, we describe the cVIL workflow in detail, with analysis tasks associated per workflow phase. 

\subsubsection{Bootstrap Phase}
In supervised machine learning, a minimal set of labeled instances is necessary to initialize the training process. 
During the Bootstrap Phase, users provide a seed set of labeled instances, which serve as the initial training data for the classifier.
This phase can be supported by several instance selection strategies for labeling.
A simple method is to provide users with random samples for labeling, to initialize the training process.
From a machine learning perspective, AL methods can be employed, with heuristics that improve labeling performance beyond random sampling.
Alternatively, bootstrapping could rely on iVIL methods, to enable users to select instances based on identified patterns in the data.
Such methods could rely on a scatter plot visualization of dimensionality-reduced instances, based on their feature values. 
In recent experiments, we have demonstrated that for the Bootstrap Phase, user-based instance selection can outperform AL~\cite{Bernard2018ComparingVL,cgf2018,euroVisShort2018,chegini2020}.
One main reason is that users have a strong focus on data properties important for labeling, which the heuristics of (model-based) AL methods may overlook, especially early in the process.
Important examples of data-centric properties for instance selection are \emph{coverage}, \emph{density}, or \emph{centrality}~\cite{Bernard2021ATO} to cover all anticipated classes with as few samples as possible. 
Given the importance of data properties for all phases of the workflow, Section ~\ref{sec:measures} provides an overview of property measures relevant for cVIL's implementation.
The bootstrapping phase is finished once all classes are provided with at least one labeled instance, forming the initial training data for supervised machine learning support.

\subsubsection{Class Selection Guidance}
As for instance-centric labeling, class-centric labeling requires a well-informed decision which class to focus on at a time. This is important because the class-centric focus could more easily lead to imbalanced training data. 
This imbalance can be addressed by ensuring all classes receive equal attention in the sequential labeling process. 
To mitigate this issue systematically, we introduce the Class Selection Guidance phase.

The principal idea is to select the next focus class wisely, before entering the Class-Based Labeling phase for this class.
This principle is inspired by AL and iVIL methods for instances, where either machine or human agents select the next instance, or the task is addressed leveraging human-machine collaboration. 
A conceptually similar algorithmic approach is \emph{Active Class Selection} \cite{Kottke2021ProbabilisticAL}, where a class is chosen for which a class label is requested from the labeling oracle.
Several strategies can guide this process, utilizing data and model characteristics revealed by property measures for data labeling~\cite{Bernard2021ATO} (given in brackets):
\begin{itemize}[noitemsep,topsep=1pt,parsep=1pt,partopsep=1pt]
    \item \emph{Training Data Balance}: Focus on the class with the fewest labeled instances to ensure balanced representation (\emph{balance}).
    \item \emph{Training Data Imbalance}: Comparing an observed distribution of labeled instances across classes with an expected distribution (\emph{imbalance}).
    \item \emph{Decision Boundaries}: Switch to an adjacent class to address challenges at class-decision boundaries between classes (\emph{borders, collision, separation}).
    \item \emph{Class Differences}: Select a class significantly different from the current focus to maintain label balance across the feature space (\emph{coverage}).
    \item \emph{Class Size}: Prioritize large or small classes based on their estimated size (\emph{size}).
    \item \emph{Class Uncertainty}: Prioritize classes with high remaining (\emph{uncertainty}), e.g., computed on the basis of margins, variance, or entropy.
\end{itemize}

Implementations of these strategies may balance human intuition and machine learning insights, ensuring optimal switches of class focus for labeling.
Combining multiple strategies may further lead to improved  Class Selection Guidance. 

\subsubsection{Class-Based Labeling Phase}
The actual labeling effort happens in this phase, where users identify multiple instances relevant for a focused class for efficient labeling.
First, users explore subsets of instances relevant for the focus class.
Interesting observations may include data characteristics of these instances, and their relationship to other classes nearby.
Users may also assess how unlabeled instances relate to those already labeled for the focused class.
For labeling, users can identify a smaller subset of instances to be labeled next. 
This subset can be selected through interactive visual exploration or by ranking instances based on properties such as \emph{class relevance}, \emph{density}, \emph{class borders}~\cite{Bernard2021ATO}. 
Users may investigate these instances in detail, apply selection and filtering, and label an identified subset with the focused class label.
Users can also adopt an approach where they exclude instances that do not belong to the focused class, removing them from the labeling scope.
These actions are repeated until users are satisfied with the class-based labeling and are ready to proceed to the next class using Class Selection Guidance.

Combined, the Class-Based Labeling and Class Selection Guidance phases continue iteratively until users have visited all classes, labeled a significant proportion of the data, and achieved a balanced training dataset, to mitigate biases and form the basis for robust model building. 
Most of the data is typically labeled during phases 2 and 3. Instances that are difficult to label are addressed in the Residuals Labeling Phase.

\subsubsection{Residuals Labeling Phase}
The Residual Labeling phase focuses on addressing remaining challenges with yet unlabeled instances~\cite{euroVisShort2018}. 
This phase transitions from a broader, class-based perspective to a detailed, instance-based focus, often dealing with outliers or ambiguous cases less representative of the overall class structure. 
Key labeling strategies observed in this phase include the following property measures for data labeling~\cite{Bernard2021ATO}:
\begin{itemize}[noitemsep,topsep=1pt,parsep=1pt,partopsep=1pt]
    \item \emph{Data Coverage}: Examining localized, previously unexplored structures within the data (\emph{coverage}).
    \item \emph{Class Separation}: Refining classes that are not yet well-separated (\emph{separation}).
    \item \emph{Class Collision}: Addressing regions where multiple classes overlap (\emph{collision}).
    \item \emph{Outlierness}: Labeling outlier instances to improve class representation (\emph{outlierness}).
\end{itemize}

At this stage, classifiers applied to the training data typically achieve a high level of performance. 
Further labeling of residuals focuses on edge cases, typically yielding only marginal performance improvements. 

The focus on atypical instances can occasionally degrade performance by introducing biases into the statistical model, reducing its generalizability. 
Despite these risks, the Residual Labeling phase is essential for comprehensive labeling, ensuring the dataset captures the full variability of each class while resolving difficult cases.
The Residual Labeling phase concludes the cVIL workflow when the user decides that all instances seem to have their correct label.


\begin{table*}
\centering
\begin{tblr}{
  width = \linewidth,
  colspec = {Q[125]Q[200]Q[675]},
  row{1-10} = {font=\small},
  column{1} = {c},
  column{2} = {c},
  cell{3}{1} = {r=3}{},
  cell{6}{1} = {r=3}{},
  vline{2} = {2-9}{},
  hline{2-3,6,9,10} = {-}{},
  hline{4-5,7-8} = {2-3}{},
}
Phase                    & Task                                & Description                                                                                                                                                                                                                                                         \\
Bootstrap                & Seed Instance Assignment            & Enable users to assign an initial set of representative instances with class labels using criteria like \textit{coverage}, \textit{density}, \textit{centrality}, or a combination. Provide algorithmic support like AL or clustering to suggest optimal seed instances based on these properties. \\
Class Selection Guidance & Labeling Status Display             & Enable users to gain an overview of the current labeling status, including progress and class-specific details.                                                                                                                                                     \\
                         & Training Data Class Balance         & Provide guidance to maintain \textit{balance} across classes in the training data, ensuring no class is underrepresented.                                                                                                                                                    \\
                         & Class Focus Guidance                & Provide further guidance to recommend the next class to focus on, leveraging strategies informed by property measures such as training data class \textit{size}, \textit{imbalance,  coverage}, \textit{separation}, or \textit{collision}.                                                 \\
Class-Based Labeling     & Instance Selection Guidance         & Provide guidance to support users in selecting instances for class-based labeling, using strategies based on class \textit{relevance}, \textit{proximity}, \textit{centrality}, class \textit{borders}, \textit{density}, \textit{disagreement}, or \textit{uncertainty}.                                    \\
                         & Multi-Instance Labeling             & Enable users to select and label multiple instances at once, based on instance similarity or other data properties. Allow users to deselect instances that do not match the focus class and efficiently execute labeling for the remaining matching instances.      \\
                         & Class Confusion Assessment          & Enable users to evaluate interactions or confusions between the current class and spatially close neighboring classes, aiding in the resolution of ambiguities based on \textit{border}, \textit{separation}, or \textit{collision}.                                                                                                     \\
Residuals Labeling       & Outlier Identification and Labeling & Enable users to identify and label atypical instances or outliers. Provide effective interactions to support users in labeling ambiguous instances of other classes, such as those at decision boundaries (\textit{border}) or overlapping regions between classes (\textit{collision}).                   
\end{tblr}
\label{table:tasks}
\caption{\new{Detailed descriptions of user analysis tasks associated with different workﬂow phases and how property measures (in \textit{italic}) support these tasks.}}
\end{table*}

\subsection{Property Measures} \label{sec:measures}

By observing participants during an instance-centric visual labeling process, we identified that they employ specific strategies while labeling~\cite{Bernard2018ComparingVL}. 
By reflecting on human-based strategies and AL-based heuristics, we further formalized these into the concept of property measures~\cite{Bernard2021ATO}. Property measures capture diverse strategies for instance selection, systematically addressing data characteristics such as density, uncertainty, and coverage. 

As outlined in the workflow description, property measures can also serve as foundational building blocks in guiding cVIL throughout various workflow phases. 
By leveraging these measures, researchers and practitioners can tackle the complexities of multi-class data labeling, helping users and incorporated guidance methods to efficiently identify relevant instances or classes.

In previous work~\cite{matt2024cVIL}, we explored three representative and complementary property measures that characterize  model outputs, data distribution, or a combination of both. 
Our experiments demonstrated how property measures effectively support class-based labeling by aligning data characteristics with specific labeling strategies. 
However, not all property measures are equally applicable to the cVIL approach. 
Certain measures align more closely with the objectives of class-based labeling, such as minimizing cognitive load, maintaining class balance, and ensuring comprehensive coverage of the data space.

In the following, we list a subset of property measures that are particularly relevant for cVIL. 
We describe these measures focusing on their implementation and applicability, to provide  guidance in realizing cVIL approaches.

\begin{itemize}[noitemsep,topsep=1pt,parsep=1pt,partopsep=1pt]
    \item \textbf{Coverage}: Assesses how well instances span the feature space, ensuring diverse representation across the data.
    Coverage is essential during the Bootstrap Phase, ensuring that initial labeled instances provide broad representation and that no regions of the feature space remain unaddressed.

    \item \textbf{Density}: Measures the concentration of instances in specific regions of the feature space. Dense areas often indicate significant or representative data regions.
    Density is highly useful in the Bootstrap Phase for identifying regions with high representational value. It also aids in the Class-Based Labeling Phase by guiding users to high-priority regions of the data.

    \item \textbf{Centrality}: Indicates the proximity of an instance to the centroid or central point of a group. It reflects the representativeness of an instance within a cluster or class.
    Centrality is highly valuable in the Bootstrap Phase, where selecting representative instances ensures robust initialization of the labeling process. It also helps refine class definitions during Class-Based Labeling Phase by focusing on prototypical examples. In cVIL we decided to use the term \emph{Eccentricity} to have small values represent prototypical items.

    \item \textbf{Imbalance}: Evaluates discrepancies in the distribution of labeled instances across classes. It compares the observed distribution with the expected one.
    Maintaining class balance is a central goal in the Class Selection Guidance Phase. Imbalance measures ensure equitable focus on all classes, preventing bias in the labeling process.

    \item \textbf{Separation}: Quantifies the distinguishability of one class from others in the feature space. It reflects how well-separated clusters or classes are from each other.
    Separation is critical in guiding the Class Selection Guidance Phase, ensuring that transitions between classes are informed by clear distinctions in the data. It helps users avoid labeling confusion and refine class structures.

    \item \textbf{Size}: Represents the count of instances within a group or class. It provides a measure of the relative representation of each class in the dataset.
    In the Class Selection Guidance Phase, size helps maintain balanced class distributions, ensuring no class is underrepresented in the training data. This supports robust model generalization.

    \item \textbf{Collision}: Indicates the degree of overlap or conflict between classes. This measure is often used to identify regions of high class collision.
    During the Class-Based Labeling Phase, this property helps users focus on regions with high inter-class confusion, enabling better resolution of overlaps and enhancing overall label quality.

    \item \textbf{Uncertainty}: Assesses the lack of confidence in instance classification, often derived from probabilistic outputs of classifiers like predictions with small margins.
    Uncertainty is highly relevant for iterative decision-making in the Class-Based Labeling Phase by prioritizing ambiguous cases for labeling.

    \item \textbf{Disagreement}: Captures the disagreement of predictions or assignments made by a model for a set of instances with high spatial proximity.
    Disagreement is particularly useful during the Class-Based Labeling Phase, where conflicting predictions can help identify challenging instances requiring human intervention. It ensures diverse perspectives in labeling decisions, enhancing dataset quality.

    \item \textbf{Border}: Measures the proximity of an instance to the edge of a cluster or class boundary. 
    This property is crucial for identifying instances at decision boundaries that require careful labeling.
    In class-based labeling, focusing on border instances ensures the resolution of ambiguities between classes. This is particularly useful during the Residual Labeling Phase, where edge cases or overlapping regions are addressed.
\end{itemize}

\new{
To summarize Section~\ref{sec:cVIL}, we introduced  the idea behind cVIL and highlighted its advantages with respect to label complexity from instance-based VIL. We further introduced the cVIL workflow with its four main phases: Bootstrap, Class Selection Guidance, Class-Based Labeling, and Residual Labeling. We identified key tasks the user must accomplish in conjunction with the interface in each phase of the cVIL workflow. Lastly, we described a subset of property measures that are particularly useful within the context of cVIL and open up a design space for cVIL approaches.
}

\begin{figure}[t]
    \centering
    \includegraphics[width=\linewidth]{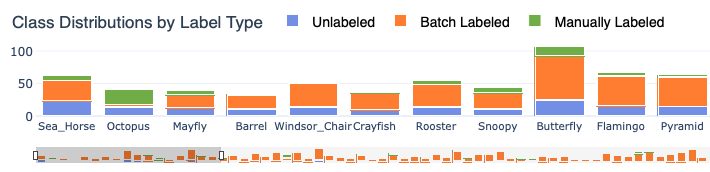}
    \caption{ The Class Label View uses a stacked bar chart to assess the distribution of instances per class and label types, color-coded as Unlabeled (blue), Batch Labeled (orange), and Manually Labeled (green). Below, a minimized version of the bar chart shows the distribution for all classes (scalable for many dozens), with the current subset highlighted with gray area. In the example, the Butterfly class stands out with high instance count, while other classes have comparatively smaller sizes. The within-class assessments reveal differing proportions of label types; for example, the Octopus class has very few batch-labeled instances -- something the user may want to address next.}
    \label{fig:cVIL_labeling_status}
\end{figure}

\section{cVIL Interface} \label{sec:interface}
We present a visual analytics approach that implements the class-based cVIL workflow for the effective labeling of large numbers of instances. Our goal was to elicit simple and well-known visualization and interaction idioms that support the cVIL workflow well. An overview of the interface can be seen in Figure \ref{fig:overview_fig}. We employ established techniques to create an intuitive and understandable interface that is also highly effective for the cVIL workflow, as described in the previous section.

\subsection{cVIL Overview}
The interface consists of four key components: Class Label View, Instance Property View, Instance Similarity View, and Instance Labeling View. 
Each component has a distinct purpose in facilitating a scalable, iterative and human-centered labeling workflow. 
The composition of all components aims at addressing challenges related to scalability and labeling efficiency while upholding the explainability of the tool.

Key design decisions include the use of color-coded stacked bar charts for quick class distribution assessment, a kernel density estimation (KDE) plot for exploring property distributions, and a scatter plot for visualizing instance relationships and addressing class overlaps. The blue, orange and green colors used in bar charts are also used in the scatter plot which enables a quick overview of instances that are unlabeled, manually and batch labeled. It is important to note that unlabeled instances are determined by model predictions rather than ground truth labels. Users can click on the bar chart to choose a focus class and further hover over a specific area in the KDE plot to choose a subclass of instances. Users can use the lasso interaction in the scatter plot for a more targeted selection of instances based on spatial proximity or visual patterns. The sample view supports batch labeling and individual instance manipulation. These components allow users to focus on both class-level imbalances and instance-level nuances.

\begin{figure}
    \centering
    \includegraphics[width=\linewidth]{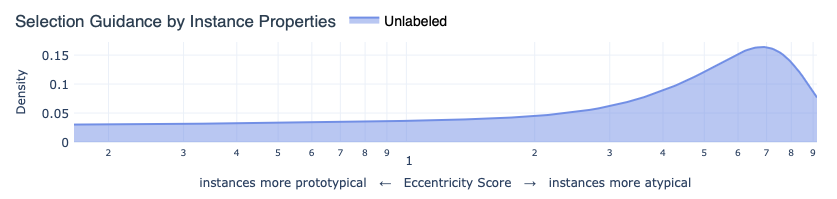}
    \caption{ The Instance Property View shows a kernel density estimation of the property measure output to visualize the distribution of values. We chose the valance of property measures such that instances with small values are more likely to be prototypical instances of the class and instances with larger values are likely to be more atypical, allowing users to quickly gauge the system's performance. The x-axis is log-transformed, allowing the visualization to show a large range of values. In this instance, the Eccentricity score is visualized, which is very skewed, which indicates that most remaining unlabeled samples for this class are distributed unevenly around the already labeled points. }
    \label{fig:cVIL_KDE}
\end{figure}

\subsection{Class Label View}

At the center of the interface is the Class Label View consisting of a stacked bar chart, as can be seen in Figure \ref{fig:cVIL_labeling_status}. Below the main stacked bar chart, there is a range slider to adjust the data range being displayed, leading to a filter for most relevant instances. Users can gain an overview of the \emph{Labeling Status} due to the color-encoding stacked bars that display the distribution of unlabeled (blue), batch labeled (orange), and manually labeled (green) instances for each class. 
This view allows users to assess class-wise imbalances at a glance. 
By clicking on a specific bar in the main focus bar chart, users can select a focus class for labeling, enabling a class-centric focus rather than instance-centric exploration. 
By using bars to show classes, the view scales for more classes compared to color-coding of classes, outperforming traditional iVIL interfaces. 
This approach is designed to enhance efficiency while encouraging users to address underrepresented classes, which is particularly important for the \emph{Training Data Class Balance} task. 
For example, when users observe that a specific class contains twice as many unlabeled instances as another, they are motivated to prioritize labeling for the underrepresented class, promoting an equitable distribution of instances across all classes. 
Sorting helps users focus only on a few relevant classes, while others can be hidden so that even dozens to hundreds of classes could be shown in theory. 
Users can easily switch between classes by clicking at the individual bars which is a particularly important for \emph{Class Focus Guidance}. 

\subsection{Instance Property View}

The Instance Property View complements the bar chart by providing a detailed view of the focus class, visualizing the density distribution of a chosen property measure. Figure \ref{fig:cVIL_KDE} shows the visualization of the \emph{Eccentricity} property measure.
The x-axis represents property values, while the y-axis corresponds to the density of instances. 
Interactive features, such as dynamic updates to the sample view and scatter plot upon hovering, allow users to explore subsets of instances. 
Hovering over a specific area on the KDE plot offers insights into how properties are distributed within the selected class. 
KDE plot supports the choice of the most prototypical instances of the class and is important for the \emph{Instance Selection Guidance}. 
The hovering interaction updates the sample view to display images with property measure values equal to or smaller than the hovered value in the KDE plot. 
Additionally, users can refine their exploration by adjusting the values through a dropdown menu, facilitating the selection of meaningful subgroups for focused labeling efforts. 
By design, images with small property measure values should represent prototypical samples of the class. 
Therefore, the most effective way to perform batch labeling of numerous samples is by focusing on the left side of the KDE plot.
Conversely, samples with large property measure values are increasingly likely to be outliers or incorrectly predicted samples. 
These are especially important for improving the model or correcting false positives.

\subsection{Instance Similarity View}

\begin{figure}
    \centering
    \includegraphics[width=\linewidth]{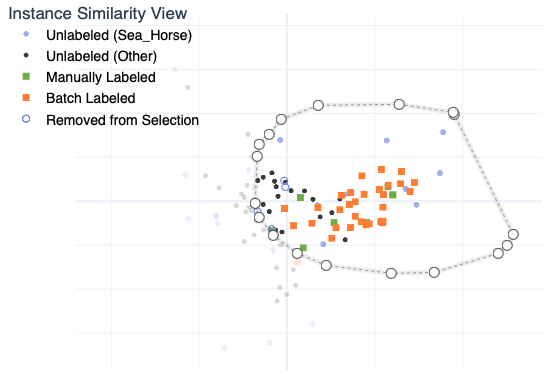}
    \caption{ The Instance Similarity View uses a scatter plot to visualize the data distribution of the Focus Class. It shows the predicted instances for the class that are still unlabeled (Blue) in relation to the already labeled instances (Green and Orange). Increasing the class rank cut-off allows for the visualization of ambiguous instances from different classes that exhibit high uncertainty regarding their class assignment to the focus class (Black). Samples that have been removed from the Instance Labeling View are encoded by a blue circle. }
    \label{fig:cVIL_instance_similarity}
\end{figure}

The scatter plot, as seen in Figure \ref{fig:cVIL_instance_similarity}, serves as an instance similarity view at an instance granularity, offering a visualization of the relationships between labeled and unlabeled instances as well as between the focused class and classes in close spatial proximity. 
Users can employ lasso interactions to select subgroups of instances based on spatial proximity or visual patterns, supporting targeted and efficient labeling workflows. 
This interaction can also support the \emph{Seed Instance Assignment}, as users can intuitively choose instances that are similar and most relevant for a focused class. 
Moreover, users can further refine their selection by applying class rank cut-offs, which help isolate ambiguous instances from various classes. By increasing the class-rank cut-off, users can visualize instances with high uncertainty regarding their classification into the focus class (black). This allows for a clearer examination of borderline cases, while samples excluded from the selection are represented by blue circles, providing a visual distinction of excluded data.
Based on the lasso selection, both the predicted samples and uncertain samples from other classes are displayed in the instance labeling view for labeling. Given the spatial proximity of the uncertain instances and correctly predicted instances in a specific region of the scatter plot, users can increase the number of correct instances in a selection for the selected class.
This enables them to prioritize instances that have lower prediction values. 
This component enhances the granularity of the labeling process, allowing users to address nuanced patterns that may not be immediately apparent from the overviews provided by the bar chart or KDE plot. 
The scatter plot also supports \emph{Class Confusion Assessment} by visualizing the spatial distribution of instances across classes, enabling users to identify areas where the current class overlaps or is spatially close to neighboring classes. 
By using lasso interactions to select instances in these overlapping regions, users can focus on ambiguous cases and resolve class confusions effectively. 
Finally, the scatter plot supports \emph{Outlier Identification and Labeling} as users can intuitively detect the spatial distribution of instances. This allows users to detect outliers or atypical points located near decision boundaries or in overlapping regions between classes in the spatial distribution of instances, and by changing the class-rank cut-offs to the lowest prediction values.

\subsection{Instance Labeling View}

\begin{figure}
    \centering
    \includegraphics[width=\linewidth]{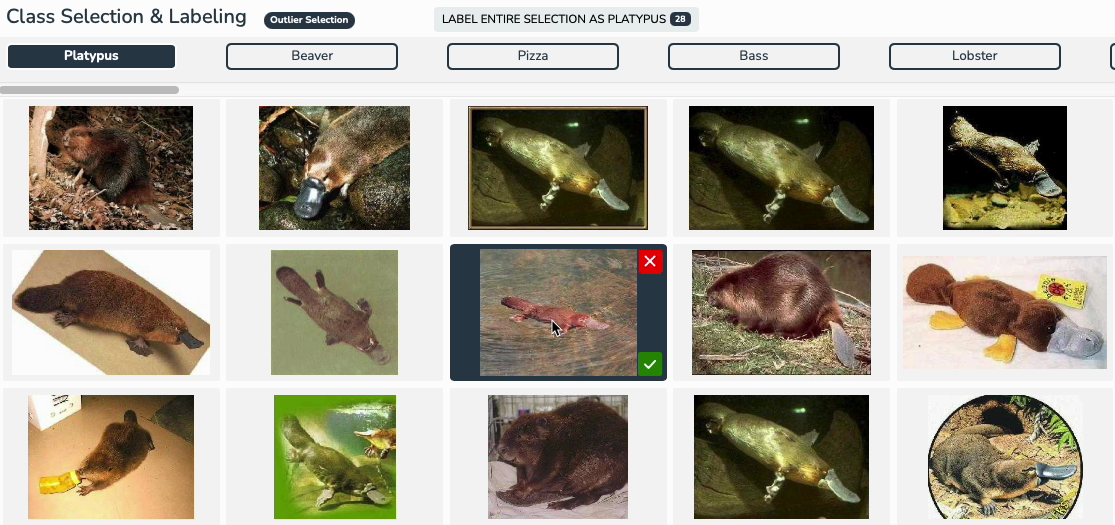}
    \caption{ The Instance Labeling View shows the actual images associated with property measures in the Instance Property View or the data points in the Instance Similarity View. It is used to examine a selection from either visualization, clean a selection, or label individual instances in the selection. When hovering over the visualizations, the Instance Labeling View displays the corresponding instances, enabling an assessment of the prediction quality in specific regions. }
    \label{fig:cVIL_labeling_view}
\end{figure}

Finally, the instance labeling view shown in Figure \ref{fig:cVIL_labeling_view} offers a preview of selected images, supporting both batch labeling and individual instance manipulation. 
Depending on the size of the window, it shows a few dozen samples corresponding to the current selection in the Instance Property View or the Instance Similarity View. 
Above the visualization, class labels are displayed as interactive buttons, with the selected label highlighted in blue. 
\emph{Single and multi-instance labeling} is supported in four ways: 1) users can batch label the instances with a single click at “Label the entire selection” where they see the number of selected images; 2) users can exclude outlier instances from the current selection by clicking the ``X'' button; 3) they can employ a drag-and-drop mechanism or confirm only selected instances to assign them to various classes; and 4) they can assign the instance to the currently focused class using the ``checkmark'' button.

This enables users to optimize the selection of images and combine instance- and class-labeling to increase efficiency depending on the visual patterns in the selection. 
For example, it might be most meaningful and efficient to first remove those instances that do not belong to a class and then label the entire remaining selection at one.

\section{Evaluation}
\edited{
The overall goal of our evaluation was to assess the scalability of the class-centric labeling approach (cVIL) with respect to two dimensions: (1) a large number of instances per class and (2) a large number of classes.
This complements the results gained in a previous experiment, where we demonstrated the potential of cVIL compared to traditional AL and iVIL~\cite{matt2024cVIL}.
To address the first dimension, we conducted a user study comparing a class-centric labeling interface (cVIL) to an instance-centric interface (iVIL) in a binary labeling task involving 2,000 images. 
This study evaluated labeling efficiency, user satisfaction, and cognitive workload, helping us understand the practical capabilities of cVIL when labeling large sets of instances.
To address the second dimension, we performed a qualitative walk-through in a usage scenario, using the cVIL prototype interface to label thousands of instances across 100 classes. 
This scenario demonstrates how cVIL scales to tasks involving many classes and highlights the workflow in detail through annotated screenshots.

For both evaluations, we employed a two-layer neural network as classifier, where the two hidden layers have 50 and 20 neurons, respectively. 
For the user study, batch-labeled instances received a lower sample weight during training. This ensured that instance-based labels remained relevant, despite the significantly larger number of samples per class as batch labeling becomes much more efficient. 
This is realized by assigning them lower costs in the loss function before back-propagation (0.1 in our experiments). 
The evaluations were conducted on an M1 MacBook Pro and used DINO~\cite{Caron2021DINO} for representation learning. 
The backbone model of DINO is a vision transformer, which was pre-trained on ImageNet \cite{Deng2009ImageNetAL}.
}

\subsection{User Study: cVIL vs.~iVIL} \label{sec:userStudy}

\edited{
The overall goal of our user study was to compare the performance, usability, and scalability of class-centric (cVIL) versus instance-centric (iVIL) labeling interfaces.
To reach this goal, we applied a within-subjects experimental design using quantitative metrics (accuracy, labeling time), cognitive workload assessment (NASA TLX~\cite{Rubio2004EvaluationOS}), and qualitative user feedback (questionnaire on preferences and usability).
The two interfaces were compared using a binary labeling task, to keep the user study within a reasonable level of complexity and time investment, allowing us to evaluate the system with a larger number of participants.
Participants labeled instances from two subsets of the CelebA dataset (glasses and hair color), each limited to 1,000 instances per class. The order of interface usage was randomized, under controlled conditions with keyboard, mouse, and external monitor.
With the binary labeling task, we could focus on class-based labeling, without class selection.
The cVIL interface displayed two KDE plots for the two classes, and used a fast min-margin criterion as single property measure.
The iVIL interface used a t-SNE~\cite{Maaten2008VisualizingDU} projection of DINO features~\cite{Caron2021DINO} in a scatter plot with instances color-coded by predicted class, as shown in Figure \ref{fig:tsne}.
}

\subsubsection{Experiment Design}
\textbf{Participants}: We had 16 participants (11 male, 4 female, 1 non-binary) with a background in computer science, recruited from a local university (4  post-graduate, 6 graduate, and 6 undergraduate level). 
Eleven participants had prior experience with machine learning. 
Participant age ranged from 22 to 40 years (median age: 26).

\textbf{Task}: The users' task was to provide labels until they thought that all instances had their correct label -- either assigned manually by the user or predicted by the model. 
The system was not initialized, which meant that users started the process with the bootstrapping phase with random instance selection for the preview. 
After selecting an initial set of few labels and initializing the model, they were asked to switch to the class-based labeling phase. 

\textbf{Data}: We used two subsets of the CelebA~\cite{celeba} dataset: one showing persons with and without glasses (\emph{CelebGlasses}) and one with people with black or gray hair (\emph{CelebHair}). 
Each class was limited to exactly 1,000 instances. 

\textbf{Independent and dependent variables}: The study employed a within-subjects design, with the interface (cVIL vs.~iVIL) as the independent variable. 
We also randomized the dataset assigned to the two interfaces, as well as the order of appearance of the interfaces. 
The dependent variables included overall accuracy (for both manual and predicted labels), labeling time, cognitive demand, and user preference. 
\new{The cognitive demand was determined by a NASA TLX questionnaire after each task and user preference was acquired through the final questionnaire after both tasks were completed by the participants and included a simple binary choice which interface was preferred as well as additional fields to describe the likes and dislikes about each interface.}

\textbf{Procedure}: For the experiment, participants used an external monitor, along with a mouse and keyboard.
Participants were first provided with a tutorial sheet that explained the system's components and how to interact with them prior to attempting the task. 
Next, the participants were asked to complete the labeling tasks. 
After solving each task, the participants completed a NASA TLX questionnaire to assess the perceived cognitive demand of the tasks. 
At the end of the study, participants were asked to express their likes and dislikes about each of the two interfaces and articulate their overall preference.

Of the 16 participants, 15 successfully completed the study. 
One user of cVIL mistakenly selected the wrong focus class for batch labeling, leading to a dramatic reduction in accuracy. 
Since this cannot easily happen when following the full cVIL workflow including focus class selection, we see this incident as non-representative and therefore excluded this user from further analysis. 

\subsubsection{Results}
\begin{figure}
\begin{minipage}[t]{0.45\linewidth}
        \centering
        \includegraphics[width=\linewidth]{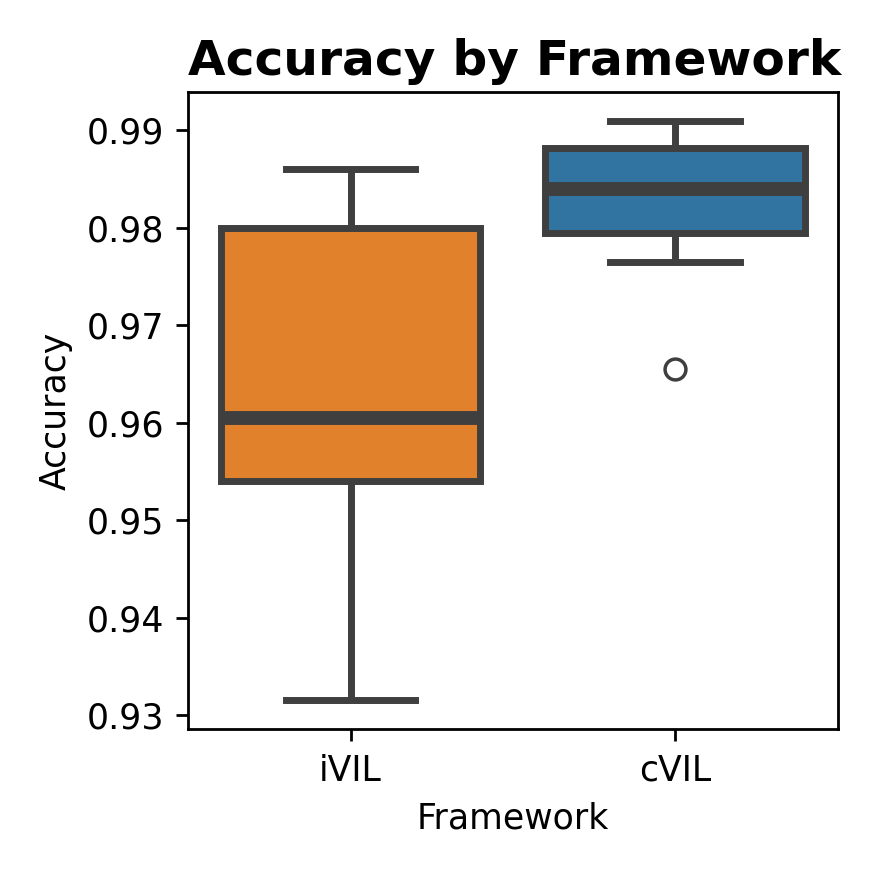}
        \caption{ Final accuracy.  }
        \label{fig:individual_results}
\end{minipage}
\hfill
\begin{minipage}[t]{0.45\linewidth}
        \centering
        \includegraphics[width=\linewidth]{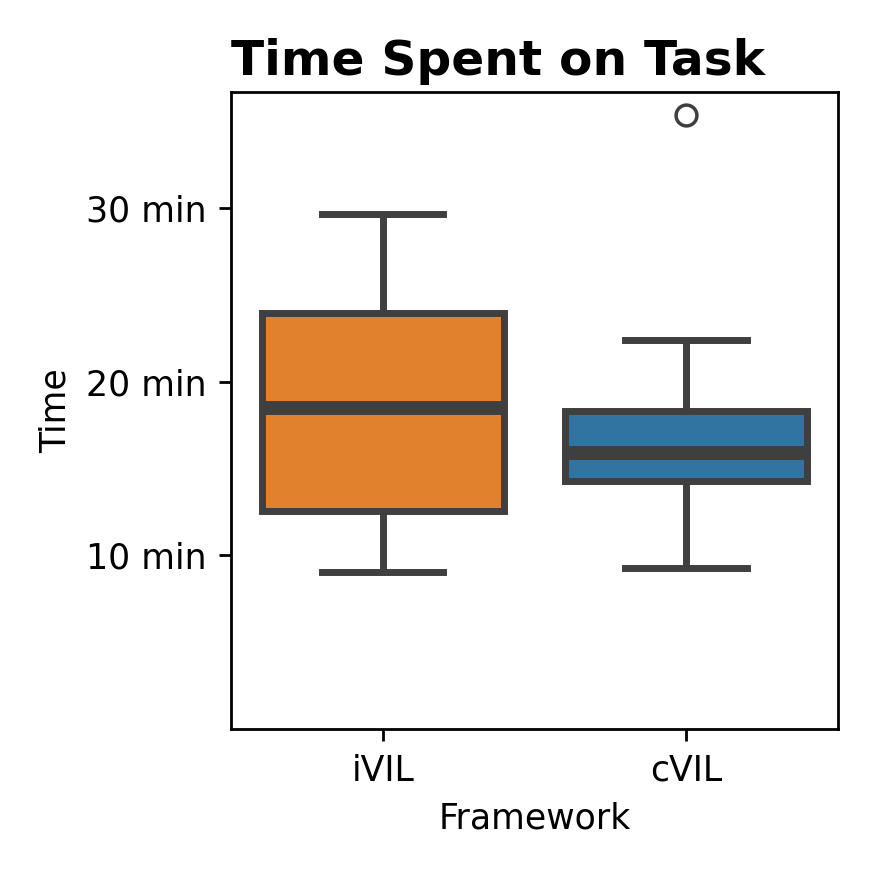}
        \caption{ Completion time. }
        \label{fig:labeling_time}
\end{minipage}%
\end{figure}

\textbf{Accuracy}: From the remaining 15 participants, all achieved a higher final accuracy using cVIL than iVIL.
The median accuracy of the exported labels compared to the ground truth was 96.05\% for iVIL, whereas it was 98.4\% for cVIL, as can be seen in Figure~\ref{fig:individual_results}, which is a statistically significant difference ($t(14)=5.784$, $p<.001$).

\begin{figure}
\begin{minipage}[t]{0.45\linewidth}
        \includegraphics[width=\linewidth]{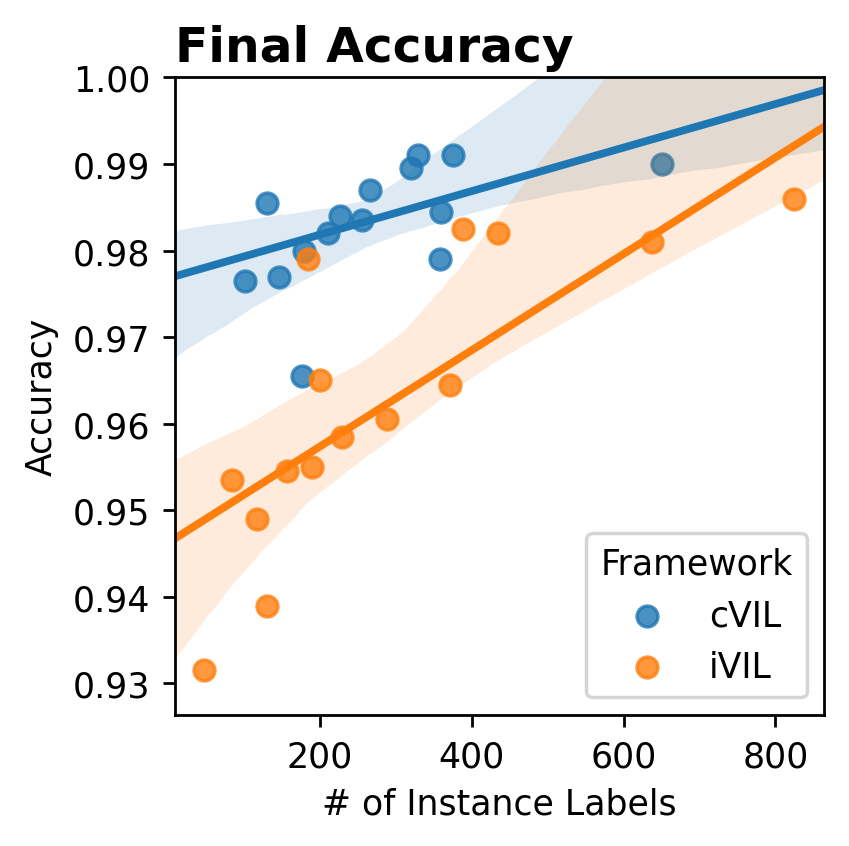}
        \caption{ Final accuracy by number of instance labels.  }
        \label{fig:lm}
\end{minipage}
\hfill
\begin{minipage}[t]{0.45\linewidth}
        \includegraphics[width=\linewidth]{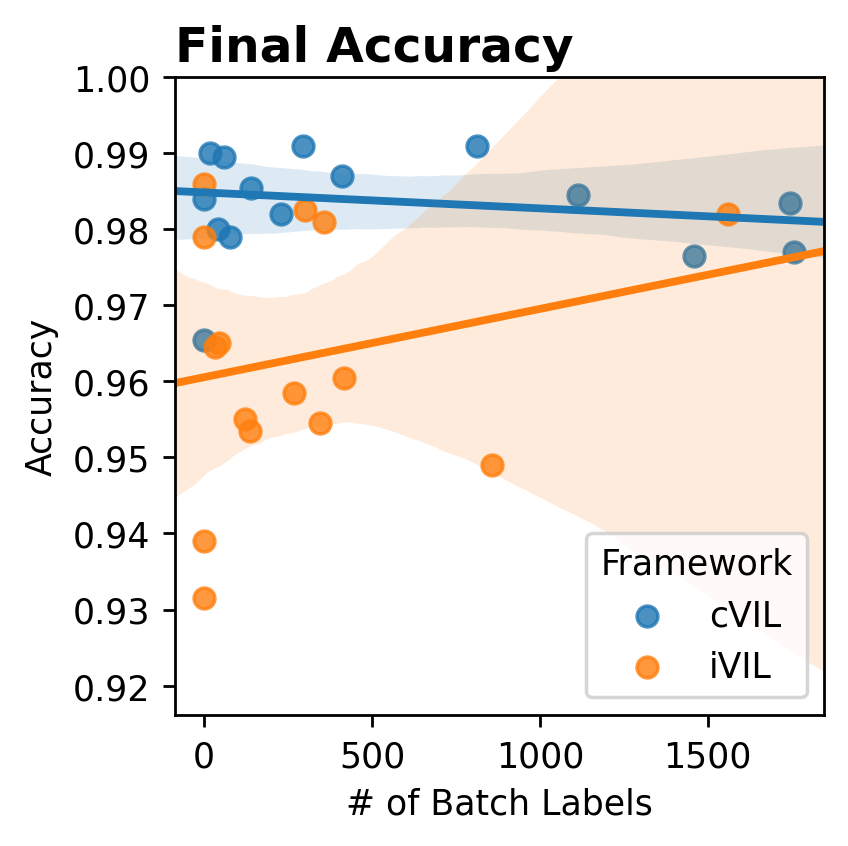}
        \caption{ Final accuracy by number of batch labels.  }
        \label{fig:lm_batch}
\end{minipage}%
\end{figure}

When comparing the final accuracy (measured against the ground truth) in dependence on the number of instance labels, we observe that cVIL achieves considerably better accuracy with fewer labels, as can be seen in Figure \ref{fig:lm}. 
The solid lines represent a robust linear regression estimation of the results for each framework. 
cVIL achieves around the same accuracy with 100 labels as iVIL with 600.
Notably, these results were achieved solely through instance labeling, as can be seen in Figure \ref{fig:lm_batch}. 
Interestingly, the accuracy is not significantly affected by the number of generated batch labels in both conditions  (except for one outlier in iVIL).
Participants batch-labeled an average of 600 samples in cVIL and 315 in iVIL, but this difference in the number of labels is not statistically significant ($t(14)=1.570$, $p=0.14$). 

\textbf{Task Completion Time}: We also measured the time it took to finish the labeling task by looking at the difference between the first and the last labeling action or model retraining. Participants generally needed less time to finish the labeling tasks in cVIL as can be seen in Figure \ref{fig:labeling_time}. 
The median labeling time for iVIL was around 18:30 minutes compared to 16:00 minutes for cVIL, however, this difference is not statistically significant ($t(14)=-1.947$, $p=0.07$).

\textbf{Task Load}: To analyze task load, we aggregated the scores from the NASA TLX questionnaire as described by Rubio et al.~\cite{Rubio2004EvaluationOS}, assigning the highest weight to Frustration, followed by Mental Demand, Effort, Performance, and finally Temporal Demand. 
No significant difference was observed between cVIL and the baseline (Wilcoxon Signed-Rank test: $z=-0.879; p=0.39$).

In the final questionnaire, however, 13 out of 16 participants expressed a preference for cVIL over the iVIL baseline. 
Participants primarily favored cVIL because they found it easier to use. 
Participants stated that the availability of uncertainty information through the min-margin property measure simplified the labeling process with cVIL and provided a better indication of the model's accuracy. 
Additionally, participants appreciated the class partitioning, which allowed them to focus on one class at a time. 
This approach required them to identify only false positives when verifying a class label, unlike the instance-based approach, which involved considering both true and false negatives.

However, participants noted a drawback: the visual representations changed minimally after model updates.
Four participants reported that the scatter plot in the iVIL baseline was also easier to understand and navigate as well as more engaging and fun to use. 
However, six other participants felt it was more tedious and ambiguous since instances were harder to find.

\edited{
The user study demonstrated that the combination of property measures and the KDE plot is effective in supporting the class-based labeling paradigm, which is a crucial component of the cVIL workflow and can support users when facing a large number of instances per class (order of one thousand instances per class). 
}

\subsection{Usage Scenario: Scalable cVIL}
We present a qualitative walk-through to demonstrate the utility cVIL.
As a complement to the user study presented in Section \ref{sec:userStudy}, this usage scenario focuses on the scalability of cVIL with respect to a high number of classes.
This usage scenario is accompanied with nine high-quality screenshots of different system states along the cVIL workflow, presented in the supplemental material.

\textbf{Dataset \& Setting}:
For the usage scenario, we examine the Caltech-101 dataset, which contains 101 diverse classes of images. To refine the dataset, the ``faces-easy'' class was removed due to the overlap with the ``faces'' class, resulting in 100 classes that have to be labeled. Each of these classes was further subsampled to a maximum of 100 instances, down from a maximum of approximately 800 for some classes, with the minimum count remaining at around 40 for some classes. This reduced the imbalance between classes to a reasonable amount and led to a total dataset size of 6,198 instances. For pre-processing, we utilized embeddings generated by the DINO model \cite{Caron2021DINO}. 

\new{
In this usage scenario, we consider Sybil, a machine learning researcher specializing in image classification tasks. 
For her research, Sybil depends on large numbers of correctly labeled high-quality data to train and validate her models. 
Her current project involves labeling a complex dataset with many classes, such as the Caltech-101 dataset, to improve the accuracy and robustness of her classification model.
}

\subsubsection{Bootstrap Phase}

\edited{
The goal during the bootstrapping phase is to manually label at least one instance for each class to initially train the model. This process is complex and tedious, often requiring significant domain expertise to distinguish between similar classes. To support the user, our approach includes several functionalities. First, class reordering, which simplifies the task by not showing already labeled classes. Second, clustering-based random sampling, which increases the diversity of instances by sampling a single instance from each cluster regardless of size. Finally, we sort classes by the least number of manual labels, which provides a clear next step and reduces cognitive load. After this bootstrapping phase, where one label per class is assigned, the model achieves an accuracy of 48\%.
}

\subsubsection{Labeling: cVIL Process}

\begin{figure}
    \centering
    \includegraphics[width=\linewidth,trim={0 21cm 24cm 12cm},clip]{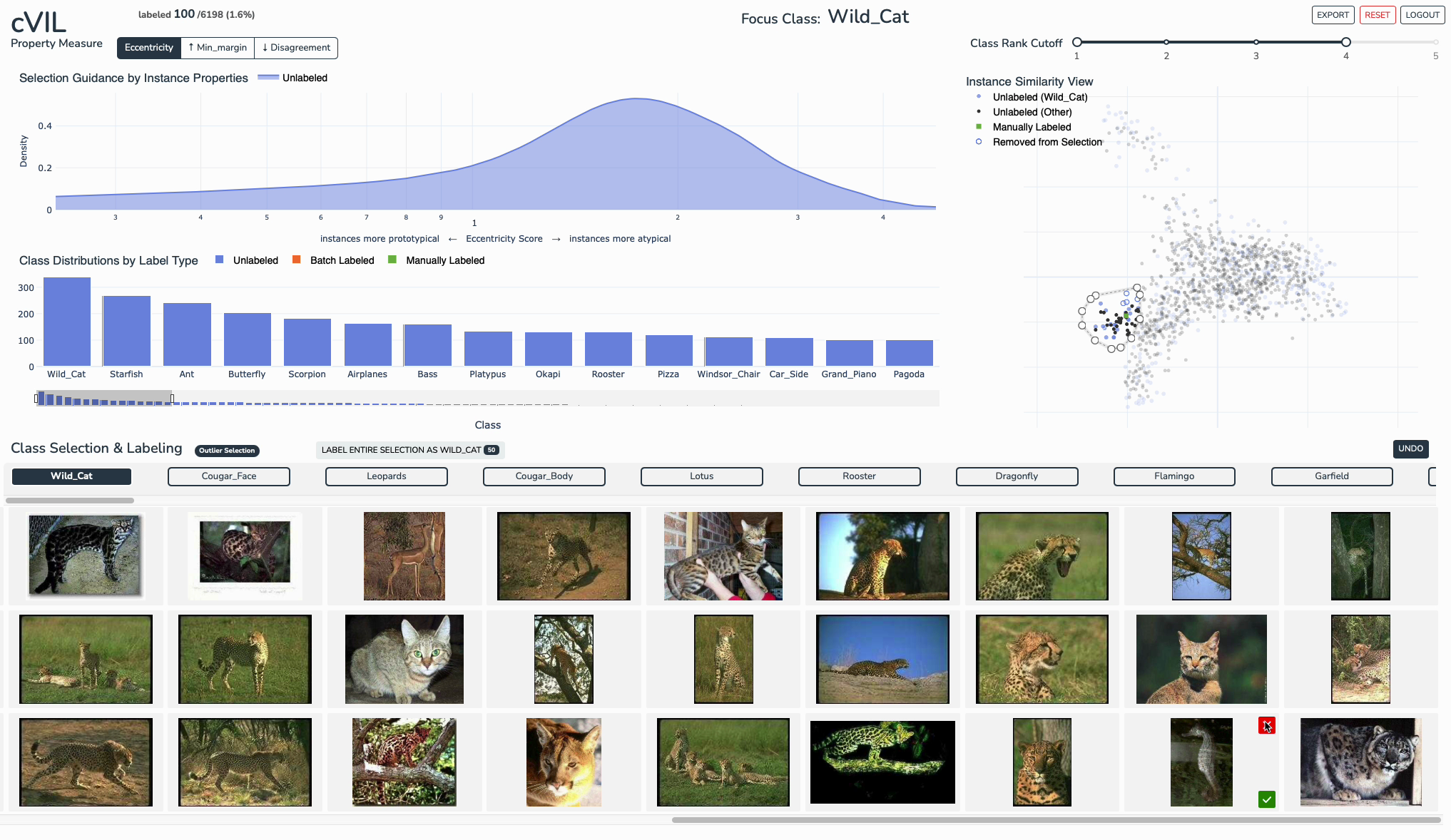}
    \caption{\new{After first training the model after the bootstrap phase, Sybil gains access to the full interface and observes in the bar chart that the class predictions are highly imbalanced. }}
    \label{fig:start_imbalanced}
\end{figure}

\edited{
After the bootstrapping phase, the visualizations (and thus the iterative cVIL process) become available (see Supplemental Figure 3). 
Sybil observes that the class predictions are highly imbalanced, with some classes having more than 200 instances assigned to them, while others have only a few, as seen in Figure \ref{fig:start_imbalanced}. The classes are ordered in ascending order based on their relative count of unlabeled samples. At this stage, all classes have exactly one label, so the ordering corresponds to the number of predicted unlabeled samples, in ascending order. Classes are ordered by the ratio of predicted unlabeled instances to labeled instances. This ordering allows users to see the current status of the labeling progress, with each class bar providing an overview of the progress for that class.

In accordance, Sybil selects the class with the most unlabeled samples according to the class ordering in order to either label a large number of samples at once or to disambiguate incorrectly predicted instances. 
Once Sybil selected a class, she can investigate the instances assigned to the focused class using the hover selection, which shows the instances Sybil hovers over in the instance labeling view. 
Sybil can now assess the focus class using either hovering the KDE plot or the scatter plot. She focuses on the KDE plot and investigates whether the prototypical instances of the class are correct by hovering over this region. Sybil decides to select instances with large and small property measure values, allowing her to visualize these regions in the scatter plot and gain insights into how the property measure relates to the spatial layout of the instances.
}

\begin{figure}[t!]
    \centering
    \includegraphics[width=\linewidth,trim={0cm 22.5cm 2cm 3cm},clip]{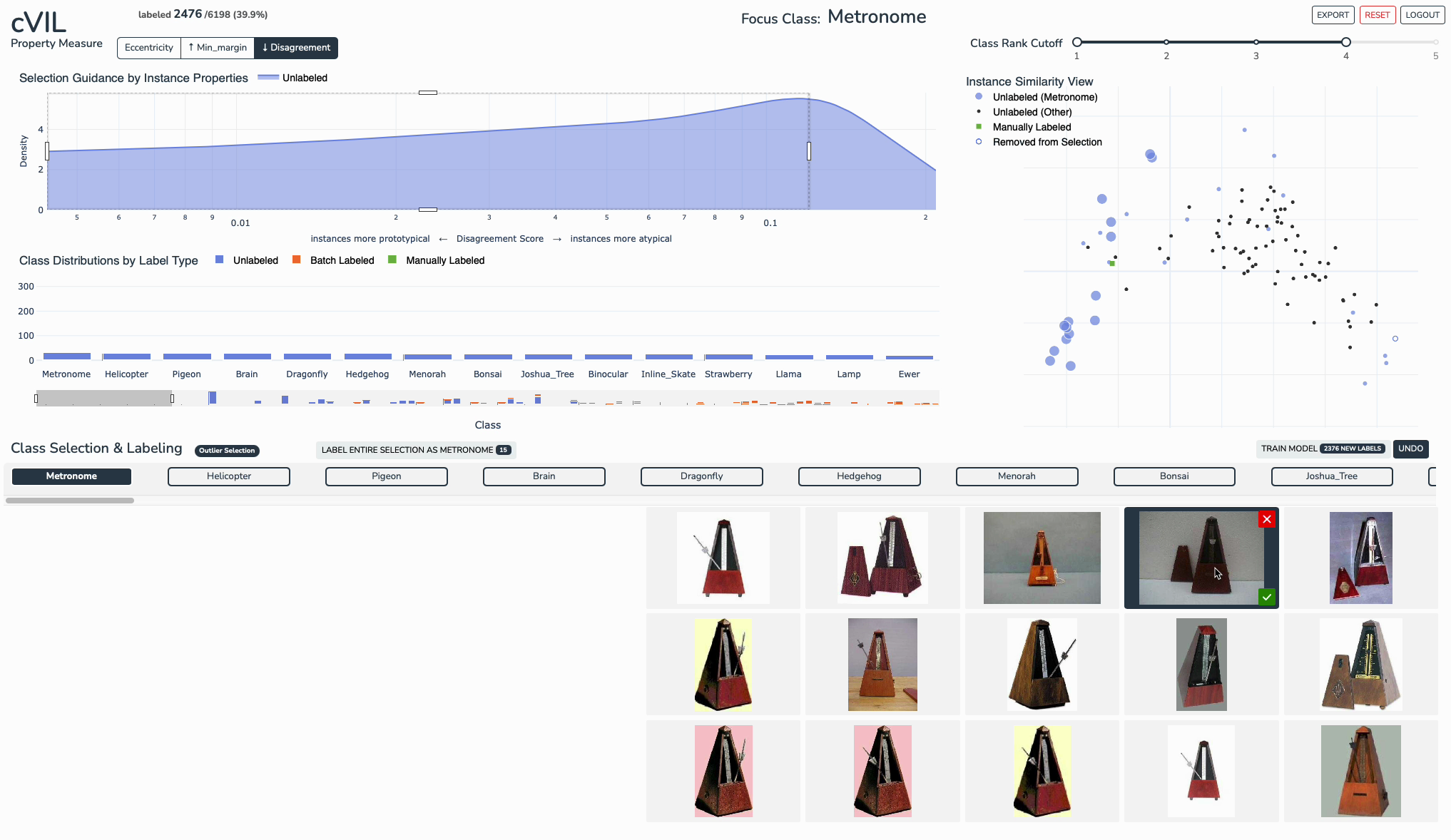}
    \caption{ For the ``Metronome'' class Sybil can still leverage the KDE plot to identify prototypical instances, even though the scatter plot is relatively spread out. }
    \label{fig:metronome_kde_selection}
\end{figure}

\edited{
After having looked at the focus class in detail, Sybil now wants to label images. She can leverage different interface features based on the prediction quality. 
When labeling the ``Grand Piano'' class with accurate predictions, she focuses on using the KDE plot to simplify the labeling process by easily identifying prototypical instances. 
She finds that the KDE plot is particularly useful for quickly selecting prototypical samples from the left side of the plot, allowing for efficient labeling of high-quality predictions. 
This also works when the predictions are not perfect, as seen in Figure \ref{fig:metronome_kde_selection}. 
If Sybil aims for more granularity, she uses the scatter plot to visualize the spatial distribution of the data points, gaining additional insights. 
By focusing on these tools, Sybil can efficiently and accurately label large subsets of the class partition with minimal effort.
}

\begin{figure}[b!]
    \centering
    \includegraphics[width=\linewidth,trim={47cm 28cm 3cm 3.5cm},clip]{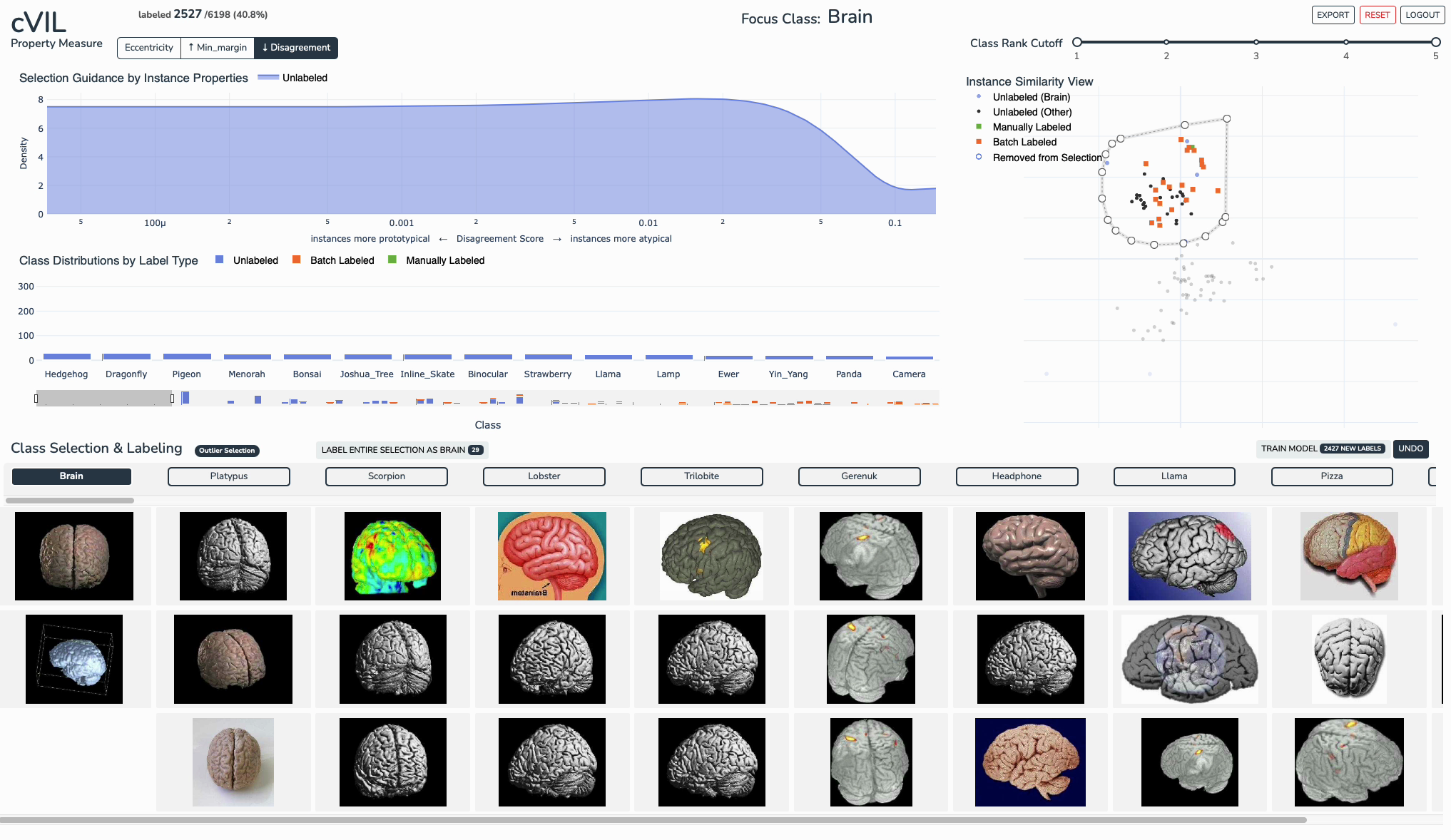}
    \caption{Sybil is directly guided to the correct cluster of instances because the already labeled instances also exclusively belong to that cluster, indicating semantic similarity. Sybil can now batch-label a larger number of instances by also increasing the class-rank cut-off, increasing label efficiency. }
    \label{fig:cluster_guided}
\end{figure}

\edited{
When dealing with a larger number of incorrect predictions, Sybil has still a lot of flexibility to improve the labels and batch label instances. 
In the scatter plot, the indication of previously labeled instances (green or orange glyphs) immediately guides Sybil to similar instances, which can be batch labeled. 
Additionally, adjusting the class rank cutoff also guides Sybil to regions with a lot of correct samples. 
For example, areas without black points indicate where the predictions are already more accurate. 
After batch-labeling a small selection, the scatter plot recomputes the layout based on the new labels, leading to better results and cleaner clusters in the data. 
For the ``Brain'' class, Sybil observes that many instances near the labeled samples are predicted to belong to other classes (indicated by black points). 
However, the proximity to the already labeled samples leads Sybil to increase the class-rank cut-off to bring in more incorrectly label samples to that region, which belong to the ``Brain'' class and can be batch-labeled effectively, which can be seen in Figure \ref{fig:cluster_guided}.
}

Finally, when dealing with classes where the number of correctly classified instances is very low, the KDE plot and scatter plot become less effective, which can be seen for the ``Gerenuk'' class. In these cases, Sybil needs to revert to instance labeling, which requires finding the correct instances manually. To do this, Sybil locates manually labeled instances. By probing the regions around these manually labeled instances, she can start to identify correctly classified instances and gradually improve the overall labeling accuracy. This process is more labor-intensive but necessary when the predictions are largely incorrect, ensuring that the users can still make meaningful progress.

With these strategies, Sybil can tackle each class individually, breaking down the problem into manageable chunks. After labeling all classes, Sybil can retrain the model. After the first iteration, the model shows significant improvement with 72\% accuracy. The labels assigned manually and through batch labeling are 97\% correct, and 2,910 instances were labeled.

\subsubsection{Residual Labeling}

\edited{
At this point, the model is already quite accurate for the prototypical samples and a large number of samples could be labeled as a result. 
Sybil now has to focus on labeling outliers and incorrectly predicted samples in each class. Batch labeling remains effective for a while, but for difficult classes, Sybil must switch to instance labeling as the predictions become increasingly unreliable. In these cases, completely unrelated images may be incorrectly predicted as the focus class, making it challenging to maintain batch labeling. For example, in the ``Butterfly'' class, the predictions are essentially random, with only one correct instance out of 22 selected samples representing 18 different classes (see Supplemental Figure 9). Despite these challenges, Sybil has successfully labeled more than 5,530 samples or almost 90\% of the data, achieving an accuracy of 97\% for the manual labels and 96.1\% for the batch labels. Batch Labels reaching almost the same accuracy as the manual labels indicates that batch labeling worked as well as individually assigning labels to each. Sybil has thus produced a high quality dataset and assigned prototypical samples to their class. If Sybil decides to continue labeling, she leaves the class-based labeling paradigm as outlined in this work and has to focus on individual instances.
}

\section{Discussion and Future Work} \label{sec:discussion}

     \paragraph{Performance and Usability} Our experiments demonstrate that class-centric visual interactive labeling can achieve superior performance compared to a purely instance-centric approach. Informal user feedback indicates a reduced cognitive load due to class partitioning, supporting the usability of our approach and the soundness of cVIL. 
     \paragraph{Handling Imbalanced Classes} While inherently imbalanced classes pose challenges, the cVIL approach offers advantages as these classes are visually highlighted in the class overview, drawing user attention for better handling. Future work should investigate the efficacy of this approach in systematically addressing class imbalance.
     \paragraph{Class Alphabets} We designed our approach for cases where class alphabets are upfront. However, domain experts are often confronted with open set recognition or class-incremental learning problems. It is open to future research to determine whether a class-centric approach makes it harder or can even be beneficial to discover unknown classes. 
     \paragraph{Applicability of Visual Idioms} Our presented class-centric workflow is per se visualization agnostic, and we presented an example with well-known visual idioms as prototype implementation. In the future, it will be important to investigate which (alternative) visual encodings and interaction techniques work well for class-centric labeling workflows. 
     \paragraph{Batch Labeling and Cognitive Flexibility} Encouraging more efficient batch labeling remains an open challenge, particularly in datasets with significant visual or semantic variability. While it may be easy to determine whether dozens of images all show the digit 1 (MNIST dataset), it may be much more difficult to determine if all images show the same type of pathology in a complex medical scan. It is therefore important to make the size (and thereby implicitly also the number) of shown images adjustable to fit the users' needs. 
     \paragraph{Residual Challenges} In scenarios involving extreme class imbalance or rare outlier instances, instance-centric residual labeling might still be necessary, as cVIL's strengths diminish in these edge cases. This underscores the need for hybrid workflows that dynamically switch between class- and instance-centric approaches.
     \new{
     \paragraph{Non-Visual Instances} In many domains, instances to be labeled may lack trivial visual representations, such as text documents, audio clips, or tabular data. For these non-visual instances, designing effective visual encodings and interaction techniques becomes crucial. Future work should explore how to meaningfully represent these instances in a class-centric workflow, ensuring that the design choices reflect the analytical focus of the involved user groups and support efficient and accurate labeling.}

\section{Conclusion} \label{sec:conclusion}
In this paper, we presented cVIL, a Class-Centric Visual Interactive Labeling workflow, addressing the challenges of labeling large datasets with numerous instances and classes.
Traditional instance-centric labeling approaches often struggle with scalability and user cognitive load, particularly in datasets with complex structures. 
By shifting the paradigm from instance-centric to class-centric labeling, cVIL reduces labeling effort, enhances efficiency, and improves user experience in managing large-scale labeling tasks.
Our work contributes to solving the scalability and usability challenges in interactive data labeling by introducing a novel workflow and interface grounded in the class-centric paradigm. 

We designed and implemented a VA prototype for cVIL and  evaluated it in two complementary experiments: a user study assessing labeling efficiency and user satisfaction in a binary labeling task with a large number of instances, and a walkthrough of the cVIL prototype to demonstrate its scalability to large numbers of classes. 
Both evaluations highlight the effectiveness of cVIL in improving labeling performance and reducing user cognitive load compared to instance-based VIL. cVIL offers scalability for large datasets, large number of classes, and adaptability to imbalanced classes. 
Its impact extends beyond traditional labeling workflows, providing a foundation for future research in class-centric labeling strategies, including scenarios like open set recognition and incremental learning. 

\section*{Acknowledgments}
This research was funded in whole or in part by the Austrian Science Fund (FWF) \href{https://doi.org/10.55776/P36453}{10.55776/P36453}. This paper was further funded by the Austrian Promotion Agency (FFG) under project grants: 898085 (TrustAI) and FO999904624
(FairAI).

%

\bibliographystyle{elsarticle-num} 
\bibliography{paper}



\end{document}


\begin{frontmatter}



\title{Scalable Class-Centric Visual Interactive Labeling -- Supplemental Material}


\author[tuw]{Matthias Matt \corref{cor1}}
\cortext[cor1]{Corresponding Author}
\ead{matthias.matt@tuwien.ac.at}
\author[uz,dsi]{Jana Sedlakova}
\author[uz,dsi]{Jürgen Bernard}
\author[fhs]{Matthias Zeppelzauer}
\author[tuw]{Manuela Waldner}

\affiliation[tuw]{organization={Institute of Visual Computing \& Human-Centered Technology, TU Wien},
            city={Vienna},
            postcode={1040}, 
            country={Austria}}
\affiliation[fhs]{organization={Institute of Creative Media Technologies, St. Pölten University of Applied Sciences},
            city={St. Pölten},
            postcode={3100}, 
            country={Austria}}
\affiliation[uz]{organization={Department of Informatics, University of Zurich},
            city={Zurich},
            postcode={8006}, 
            country={Switzerland}}
\affiliation[dsi]{organization={Digital Society Initiative, University of Zurich},
            city={Zurich},
            postcode={8006}, 
            country={Switzerland}}
\begin{abstract}
This document contains the supplemental materials that accompany our cVIL class-based labeling approach.
\end{abstract}







\end{frontmatter}



\section{Usage Scenario - Flip Book with Significant Stages}

\clearpage

\begin{figure}[t]
    \centering
    \includegraphics[width=\linewidth]{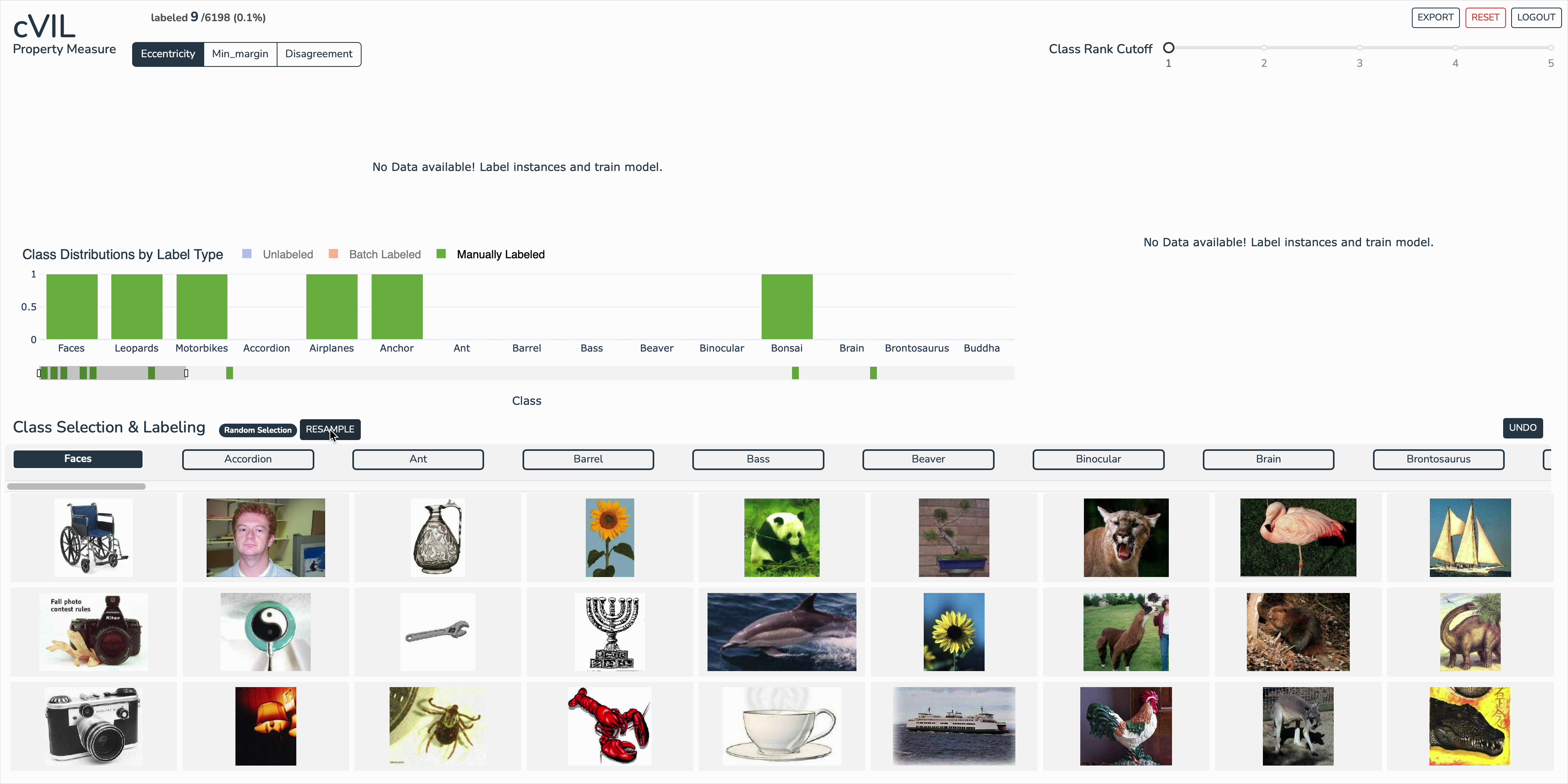}
    \caption{First, the user has to initialize the system, which is implemented as a random selection. The user can either try to match the images to labels or the labels to images. In this case, the user wants to find an instance of accordion to label this class. Since the current random selection does not contain an accordion, they decide to resample the selection until one is being sampled. After labeling and image, the class has a label and the classes in the labeling view get reordered. The classes with existing labels now appear last in the list. }
    \label{fig:cold_start}
\end{figure}

\begin{figure}
    \centering
    \includegraphics[width=\linewidth]{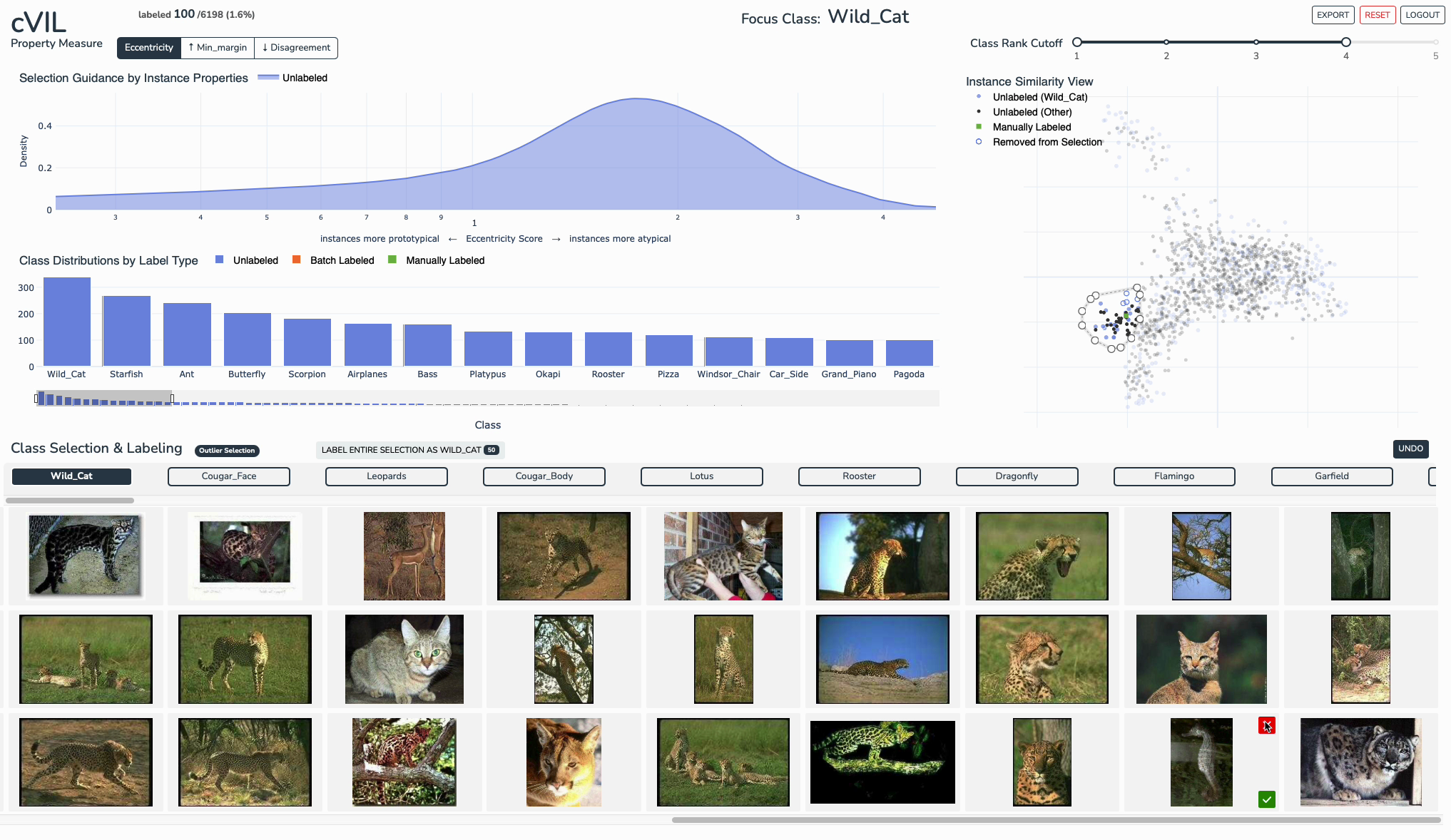}
    \caption{This is the state just after bootstrapping with a single label for each class and 100 labels in total. The user starts with the class that has the most unlabeled samples, which is \enquote{Wild\_Cat} in this case. Due to seeing many incorrectly predicted labels the KDE plot the user switched to the scatterplot as there are too many incorrectly predicted labels. In the scatterplot, the user found a small region that contains wild cats. Increasing the class rank cutoff, similar samples from other classes are mapped to the same region. With this selection, the user cleans the selection by removing incorrect predictions and could label 19 instances as wild cats. Alternatively, the user could have labeled the instances using the label suggestions for the selection, which included most relevant classes. }
    \label{fig:wild_cat_start_bad_predictions}
\end{figure}

\begin{figure}
    \centering
    \includegraphics[width=\linewidth]{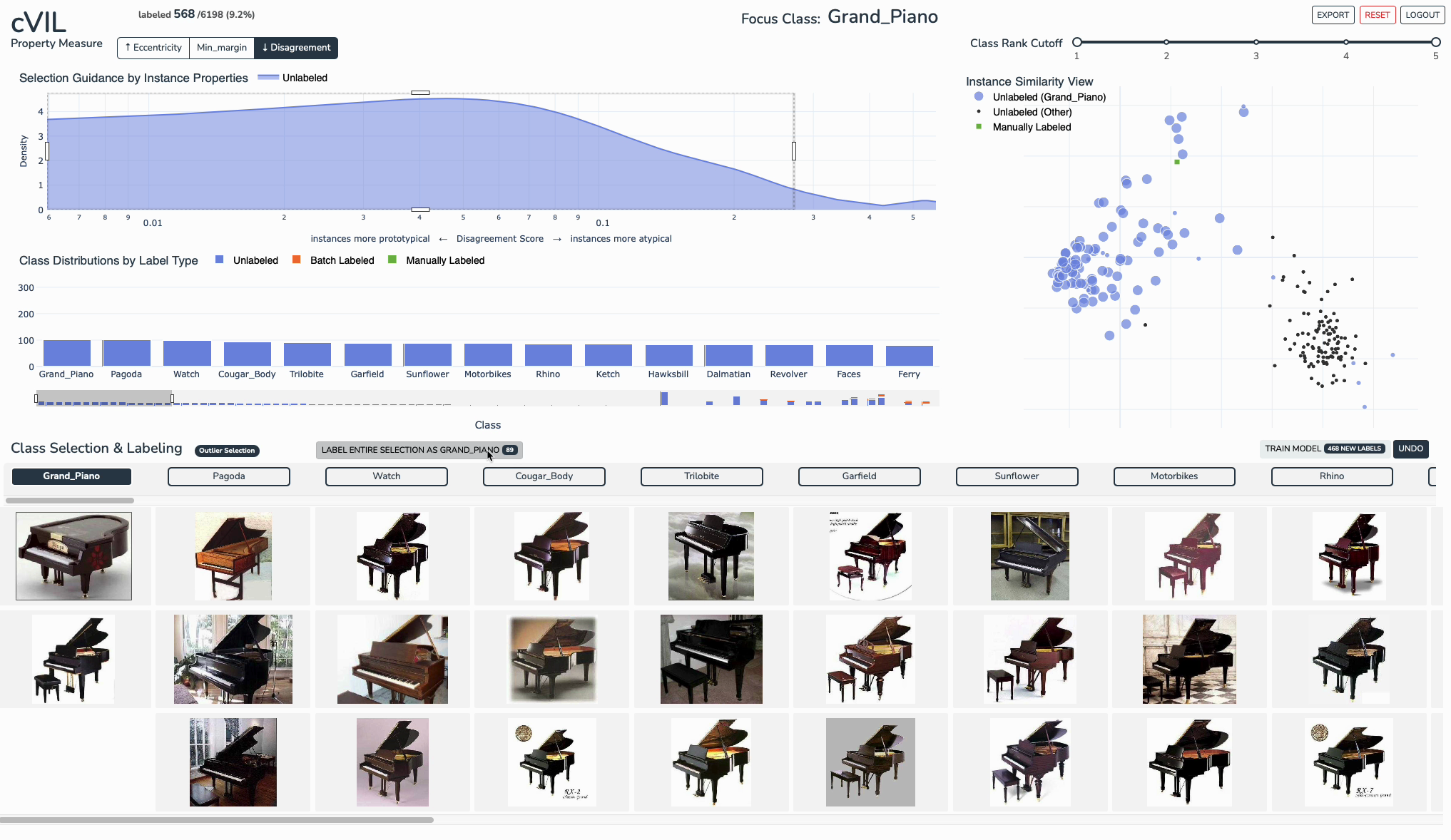}
    \caption{Continuing with the labeling process, the user also encounters classes with good predictions such as \enquote{Grand\_Piano}. Here, the KDE plot is a very simple tool to quickly find a large selection without looking at the data distribution. The user can quickly confirm that the labels are correct by hovering over the KDE plot. The user makes a selection of the instances in the KDE plot, which are also highlighted by larger glyphs in the scatter plot. We can see the instances are all part of one large cluster. Confirming that the instances in the selection are correct the user can label 89 instances at once and continue labeling a different class. }
    \label{fig:grand_piano_easy}
\end{figure}

\begin{figure}
    \centering
    \includegraphics[width=\linewidth]{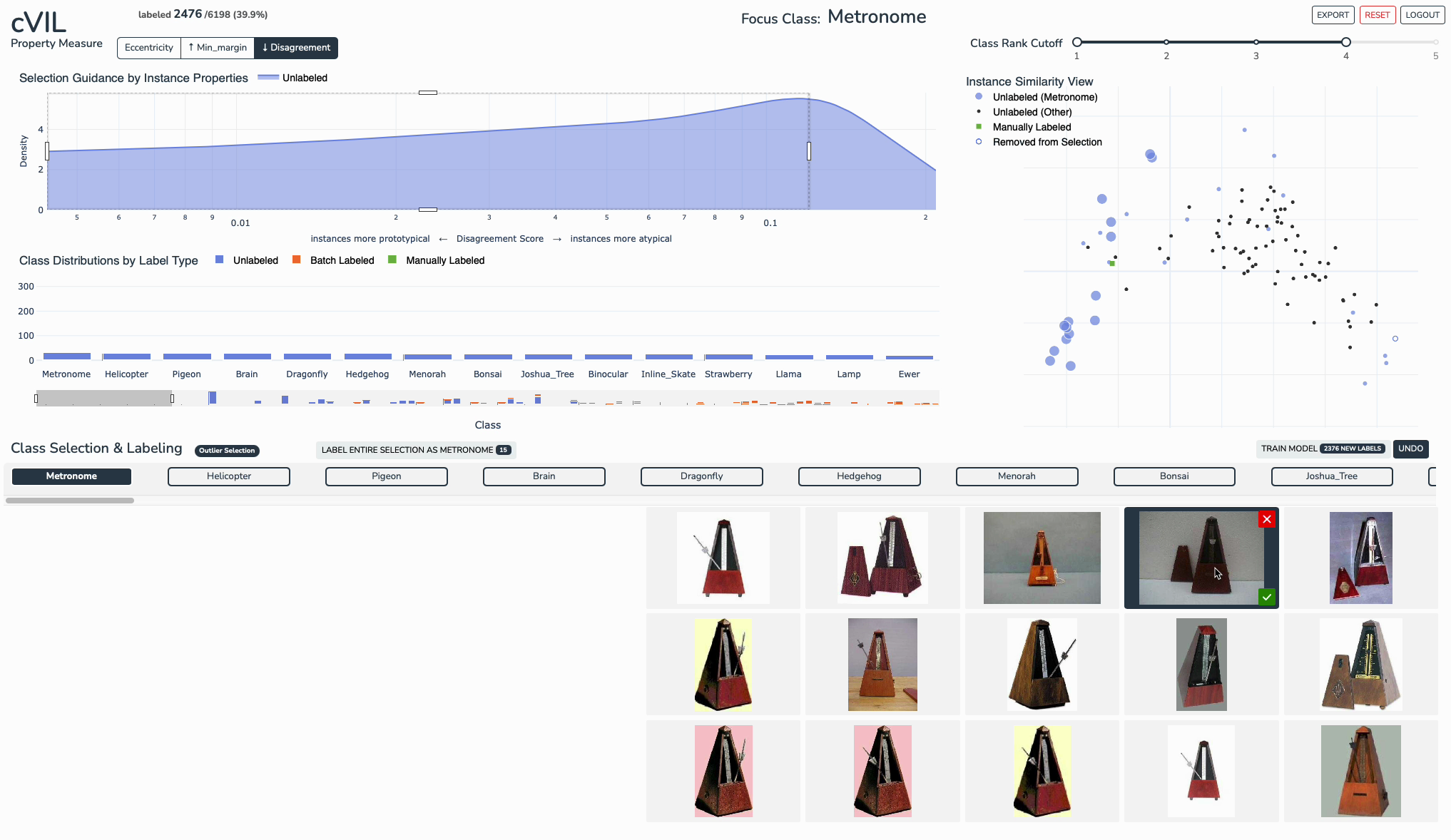}
    \caption{The user encounters more classes with imperfect predictions. For the \enquote{Mentronome} class the KDE plot was still helpful to find an initial selection. The user made an initial selection in the KDE plot. As we can see, this selection consists of a cluster on the right in the scatter plot. One instance in the selection was incorrect, which was subsequently removed. This incorrect instance was a small Buddha statue, which was not part of the main cluster as we can see by the rightmost point in the scatter plot, indicating the removed instance. The user can now label the selection. }
    \label{fig:metronome_before}
\end{figure}

\begin{figure}
    \centering
    \includegraphics[width=\linewidth]{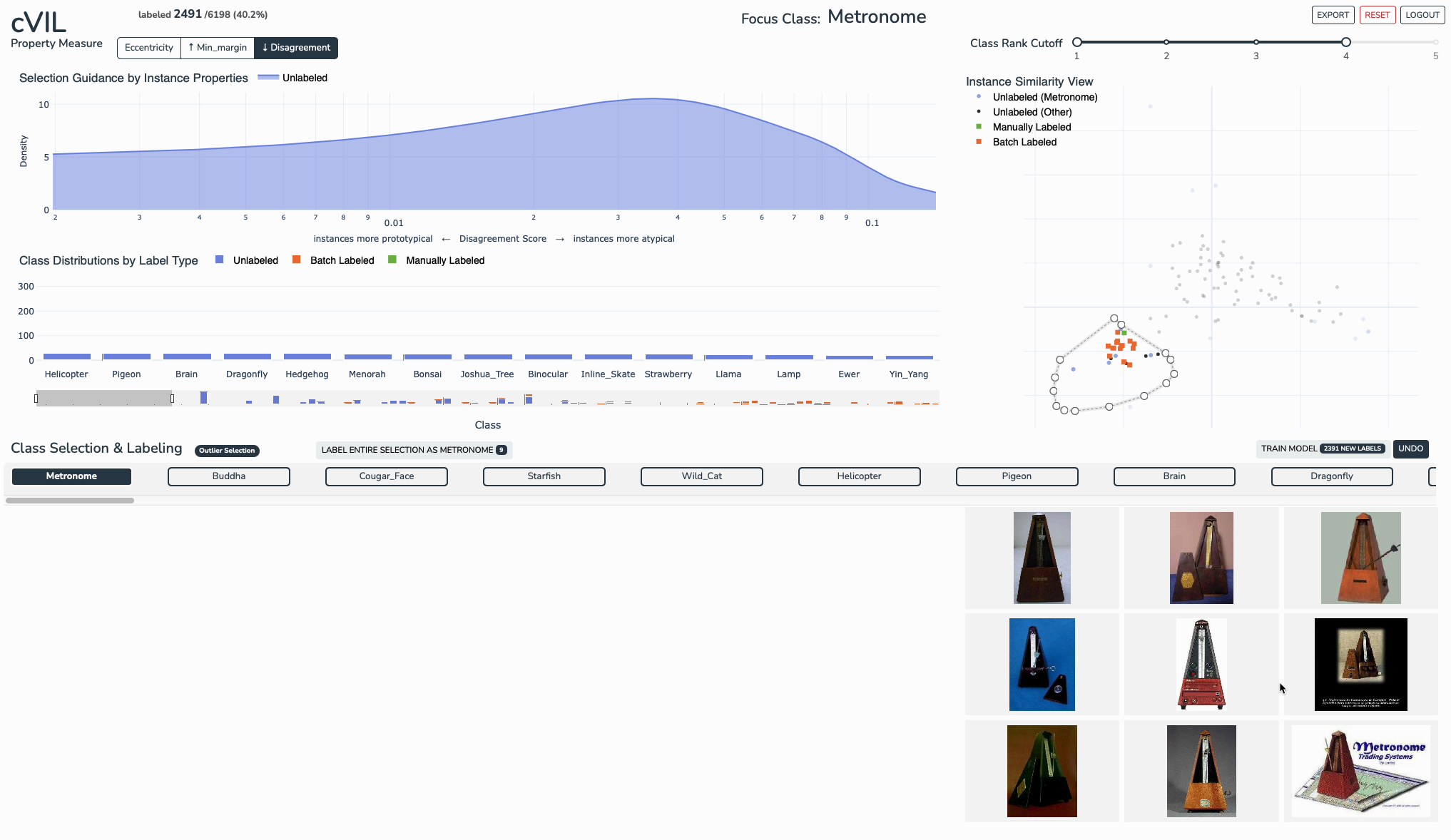}
    \caption{Continuing with this class, the scatter plot projection is recomputed with the new batch labels that were added before. This changed the layout of the instances with one cluster clearly emerging. The orange glyphs for batch labeled instances further highlight this cluster. The user selects the cluster and finds more relevant instances belonging to the \enquote{Metronome} class. No points with other predictions (black) were mapped to this region, indicating that the precision for this class is high. After confirming that the remaining instances are incorrect using the KDE plot, the user labeled all relevant instances in the class and continues labeling a different class. }
    \label{fig:metronome_after}
\end{figure}

\begin{figure}
    \centering
    \includegraphics[width=\linewidth]{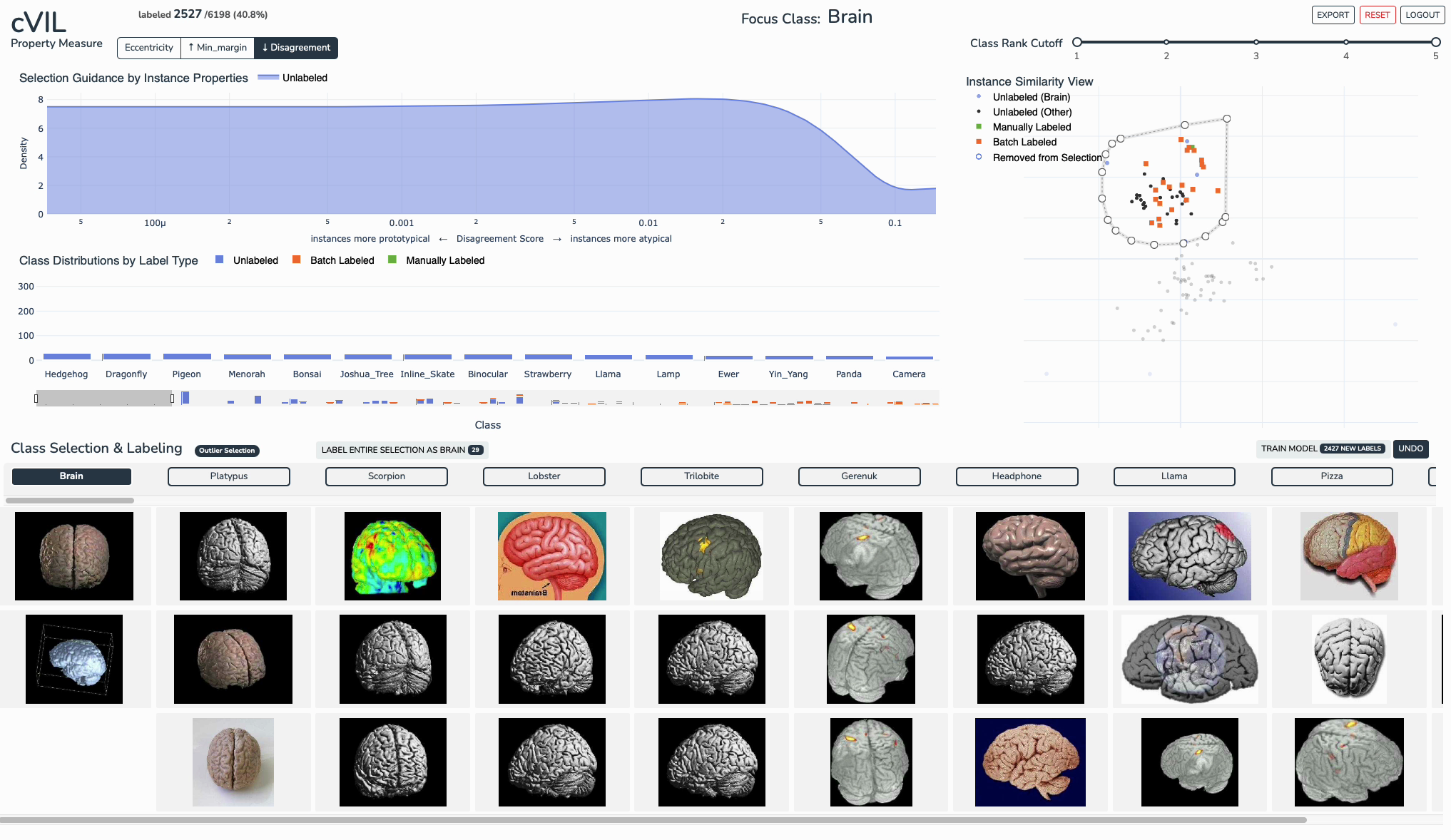}
    \caption{ For the \enquote{Brain} class, the class rank was instrumental for labeling instances. After an initial label, a clear cluster of batch labeled instances emerges. This triggers the user to increase the class rank cutoff. As we can see, many more instances from other classes are mapped to the same region as the previously batch labeled instances. With this, the user could still label 29 additional samples. This might indicate that the precision of the model for the \enquote{Brain} class was high but the recall is low. }
    \label{fig:brain_class_rank}
\end{figure}

\begin{figure}
    \centering
    \includegraphics[width=\linewidth]{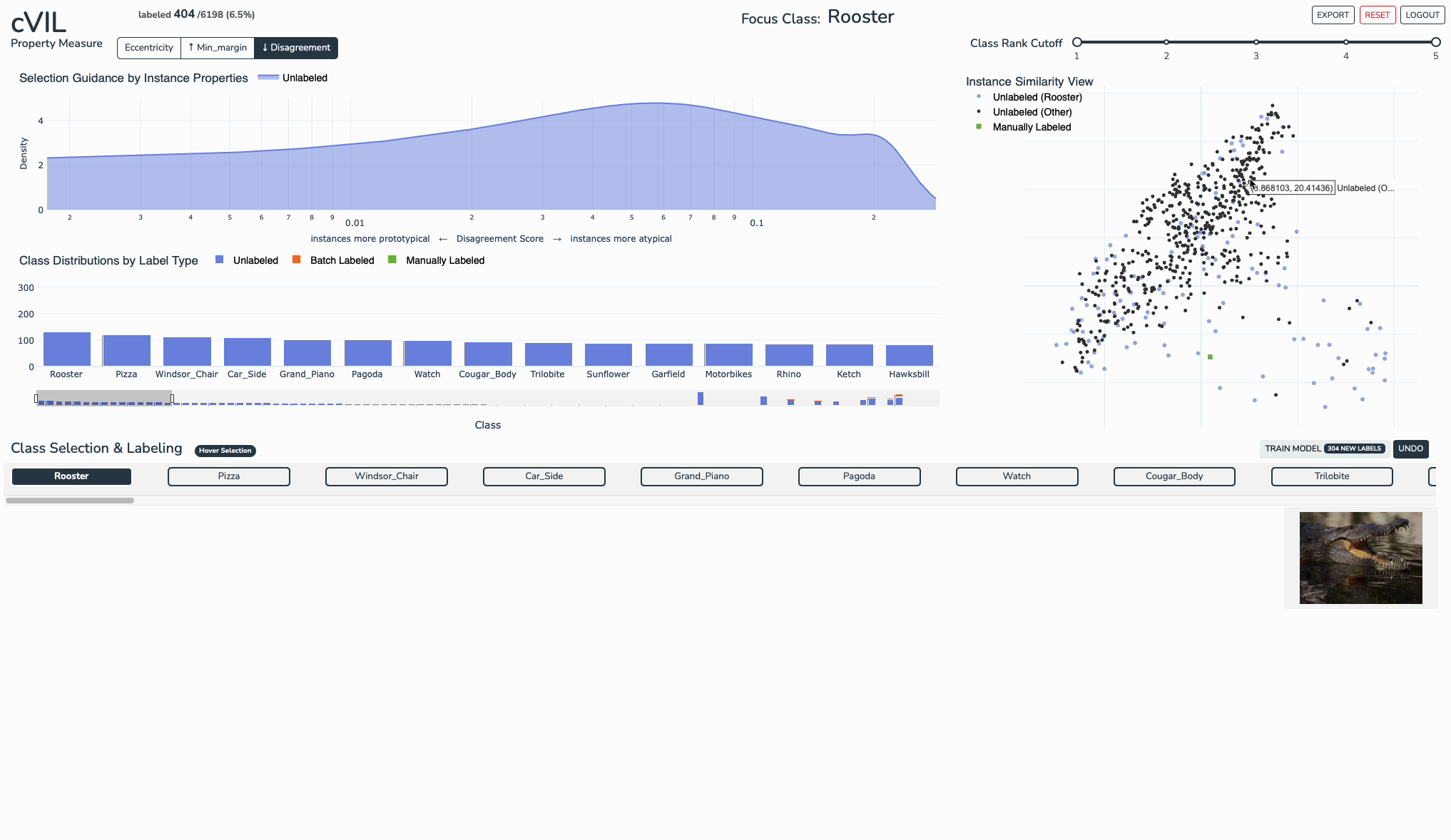}
    \caption{Another way to quickly find clusters for the selected class when the model misclassifies many instances is to increase the class rank cutoff. Here the main cluster contains almost all of the instances from other classes. The user then quickly confirms that the main cluster does not contain the target class. The user then selects the cluster where the fewest instances from other classes were added and ignores the rest. This represents the oppoosite case from Figure \ref{fig:brain_class_rank} where in this case the model potentially has low precision but high recall. }
    \label{fig:rooster_class_rank}
\end{figure}

\begin{figure}
    \centering
    \includegraphics[width=\linewidth]{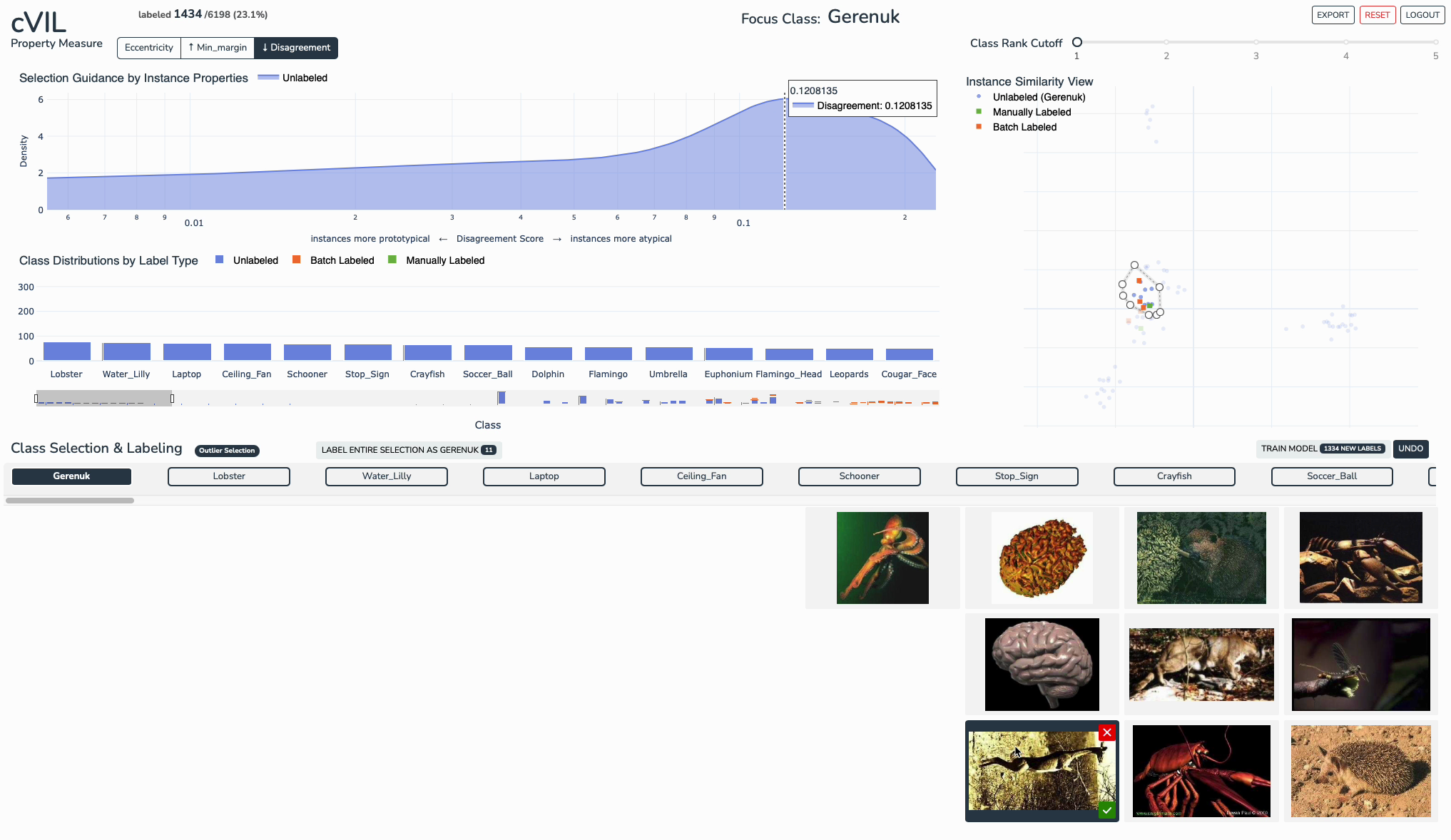}
    \caption{ When the prediction of the model are incorrect overall with very low precision and recall, the user has to rely on probing the data to find some correctly classified instances. For the \enquote{Gerenuk} class, the user saw that the KDE plot was not helpful for indentfying the correct instances and that the main cluster of the scatter plot did only contain a small number of correctly predicted instances. The user then probes regions in the scatter plot. The few correct instances can be  quickly labeled by clicking on the checkmark icon. }
    \label{fig:gerenuk_manual_probing}
\end{figure}

\begin{figure}
    \centering
    \includegraphics[width=\linewidth]{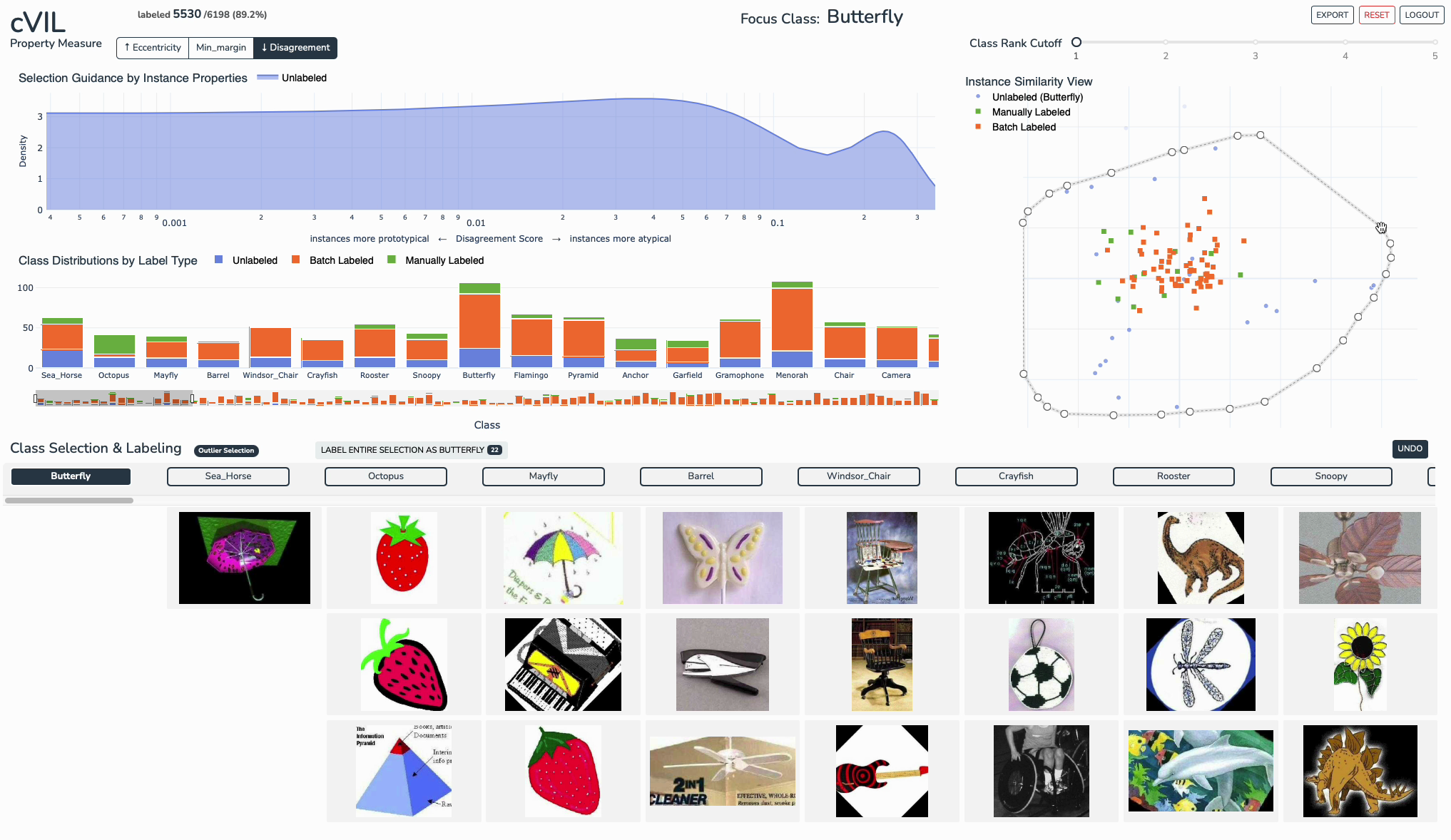}
    \caption{ When the system reaches a state where it becomes more difficult for the model to correctly assign instances, the user has to rely more on manual labeling. For the case of the \enquote{Butterfly} class, a central cluster is clearly visible, which the user then selects to label. However, the instances around the cluster are varied and almost random. The user can choose to ignore these instances for now, hoping that they get corrected while labeling the other class or manually assign each instance to its correct class. Given that almost every instance belongs to a different class, this is quite intensive. For this cases, we are in the residual labeling stage of the system where the class-based approach breaks down. }
    \label{fig:butterfly_residual_labeling}
\end{figure}